\documentclass[a4paper,11pt]{article}
\usepackage{amsmath}
\usepackage{amssymb}
\usepackage{amsfonts}
\usepackage[pdftex]{graphicx}
\usepackage[hang,small,bf]{caption}
\usepackage[subrefformat=parens]{subcaption}
\usepackage{float}

\begin{document}
\begin{titlepage}

 \begin{normalsize}
  \begin{flushright}
UT-Komaba/24-6\\ KOBE-COSMO-25-02\\ December 2024
  \end{flushright}
 \end{normalsize}

 \vspace{1cm}
 \begin{center}

  \begin{LARGE}
{\bf Stochastic quantization with discrete fictitious time}
  \end{LARGE}

\vspace{2cm}
\begin{large}
Daisuke~Kadoh~$^a$,\hspace{2mm}
Mitsuhiro~Kato~$^b$,\hspace{2mm}
Makoto~Sakamoto~$^c$ \hspace{1mm}and\hspace{1mm}
Hiroto~So~$^{a}$
\end{large}

\vspace{10mm}
\begin{it}
$^a$Institute for Mathematical Informatics, Meiji Gakuin University, Yokohama, Kanagawa 244-8539, Japan\\[4pt]
$^b$Institute of Physics, University of Tokyo, Komaba, Meguro-ku, Tokyo 153-8902, Japan\\[4pt]
$^c$Department of Physics, Kobe University, Nada-ku, Hyogo  657-8501, Japan\\[4pt]
\end{it}

\date{}

\end{center}
\vspace{4cm}

\begin{center}{\bf Abstract}\end{center}
\begin{quote}\begin{normalsize}

We present a new approach to stochastic quantization \`a la Parisi-Wu with a discrete fictitious time. 
The noise average is modified by weights, which results in the equivalence in the large time limit to the correlation function of the corresponding quantum field theory \textit{without taking any continuum limit of fictitious time}.
We test our method in a zero-dimensional toy model both perturbatively and numerically.

\end{normalsize}\end{quote}

\end{titlepage}
\vfil\eject

\section{Introduction}

The Parisi-Wu stochastic quantization \cite{Parisi:1980ys} is an interesting approach to understanding quantum theory from various aspects.%
\footnote{Refs.\,\cite{Damgaard:1987rr, Namiki_lecture} provide a reveiw of  stochastic quantization.
See those references for important early studies.}
For a given $d$-dimensional field theory, an extra continuous time is introduced and  Langevin dynamics is given by a classical motion along the fictitious time with Gaussian noises. Parisi and Wu have provided a prescription that correlation functions of $d$-dimensional quantum field theory are obtained from a noise average of equal time solutions to the Langevin equation. Roughly speaking, classical dynamics with fluctuations caused by noises behaves as a quantum field theory.

There have been many attempts to prove the stochastic quantization. The probability distribution derived from the Langevin equation satisfies a Fokker-Planck (FP) equation,
and a formal proof was given by using the fact that the FP equation yields the Boltzmann distribution of the $d$-dimensional field theory at the large time limit. 
In Refs.\cite{Parisi:1982ud}, 
supersymmetry has also been used for proofs. The Nicolai map \cite{NICOLAI1980419} allows us to write the noise average of correlation functions as an ordinary path integral form with a supersymmetric action. Proofs using supersymmetry were given in perturbation theory \cite{Nakazato:1983gm,Egorian:1983wz}, and then a non-perturbative proof has been provided \cite{Kirschner:1984sx,Gozzi:1984au}.%
\footnote{As noted in Refs.\,\cite{Kugo:1989xs,Zinn-Justin}, the supersymmetry invariance is, however, broken at the initial or final value of the fictitious time.  The proofs are based on the assumption that such a breaking effect could disappear in the equilibrium limit.  }

Stochastic quantization using the Langevin equation has rich numerical applications. 
In the field of lattice QCD \cite{wilson1974confinement, montvay1994quantum}, the Hybrid Monte Carlo method 
\cite{Duane:1987de} which is an extension of the Langevin simulation is widely used as a major computational method. 
In actual numerical applications, the $d$-dimensional action that gives the drift term is given by a lattice action, and the fictitious time is discretized. In a simple Langevin simulation, the discrete time must be taken to the continuous limit at the end of the calculation, which leads to an extra cost in numerical calculations. 
However, if we can establish a method of stochastic quantization that is valid without taking continuous limit of discrete time, it could lead to new developments in the field of numerical simulations of quantum field theory.

In this paper, we propose a method of stochastic quantization that remains valid in discrete time. 
Our proposed method modifies a noise average with weights. By taking the weight function appropriately, we show that stochastic quantization can be established even in discrete time. The proof utilizes a supersymmetric formulation. As a preparation, we give another proof of the Parisi-Wu stochastic quantization with continuous time by moving to a form of supersymmetric theory that has no supersymmetry breaking at the boundary.  We then show our results in discrete time using an essential part of that proof. By naively taking the continuous-time limit from our method, we give a non-perturbative alternative proof of the Parisi-Wu stochastic quantization. There are some subtleties in the continuous-time approach, but such subtleties can be resolved explicitly in the discrete-time approach. We also test our method in the zero-dimensional toy model using perturbation theory and numerical calculations.

The rest of this paper is organized as follows. In Section \ref{SQ}, we review the Parisi-Wu stochastic quantization method for $d$-dimensional scalar theory.
We begin with introducing the Langevin equation, after which we prove Parisi-Wu's statement from an equivalent supersymmetric theory and its deformation with exact supersymmetry at the boundary. 
Section~\ref{PWL} is the main result of our paper in which a new theorem on stochastic quantization that holds on a finite discrete time is presented.  
In Section~\ref{toy}, our method is applied to a zero-dimensional toy model both perturbatively and numerically to confirm its validity. A final section is devoted to the summary. In Appendix~\ref{A}, we refine the proof of Parisi-Wu stochastic quantization presented in Section \ref{SQ} by discretizing path integral measure.  Also, the details of the perturbative calculation are explained in Appendix~\ref{A.perturb}.

\section{Parisi-Wu stochastic quantization}
\label{SQ}

In this section, we outline the Parisi-Wu stochastic quantization, with an emphasis on reorganizing the key concepts required for our method.
The noise averages for solutions of the Langevin equation can be expressed in terms of path integrals in supersymmetric theory. 
This kind of formulation using supersymmetric theory is suitable for the discrete-time case given in next section.

\subsection{Langevin equation}

We consider a quantum field theory (QFT) of an interacting real scalar field $\varphi(x)$
on a euclidean $d$-dimensional spacetime, 
\begin{align}
S_{\textrm{QFT}}[\varphi]
=\int d^d x \, \left\{ \frac{1}{2}(\partial_\mu \varphi(x))^2 +V(\varphi(x)) \right\}.
\label{QFT_action}
\end{align}
In the path integral formulation, 
the correlation function of $n$ scalar fields is given by 
\begin{align}
\langle  \varphi(x_1)\varphi(x_2) \cdots \varphi(x_\ell)  \rangle_{\textrm{QFT}} = 
\frac{1}{Z_{\textrm{QFT}}}\int [d\varphi] \, \varphi(x_1)\varphi(x_2) \cdots \varphi(x_\ell) \, {\rm e}^{-S_{\textrm{QFT}}},
\end{align} 
where $Z_{\textrm{QFT}}=\int [d\varphi] \,  {\rm e}^{-S_{\textrm{QFT}}}$.

We introduce a $d+1$ dimensional field 
$\phi(x,t)$ for a fictitious time $t \ge 0$.  
The associated Langevin equation is defined using a Gaussian noise function $\eta(x,t)$ as
\begin{align}
\partial_t \phi(x,t)  = \square \phi(x,t) -V^\prime(\phi(x,t)) + \eta(x,t),
\label{langevin}
\end{align}
where $\square =\partial_\mu \partial_\mu$. 
The first two terms on the RHS of 
Eq.\eqref{langevin} correspond to the gradient of the action 
because $\delta S_{\textrm{QFT}}/\delta \varphi = -\square \varphi +V^\prime(\varphi)$.
The initial value of $\phi(x,t)$ at $t=0$ is given by as  a fixed function $\phi_{0}(x)$ as
\begin{align}
\phi(x,t)|_{t=0} =  \phi_0(x).
\label{initial_condition}
\end{align}
For the given noise function
$\eta$, a solution $\phi$ satisfying Eq.\eqref{langevin} with the initial condition \eqref{initial_condition} 
is denoted as $\phi_\eta$. 
We note that $\phi_\eta(x,t)$ does not depend on the noise function
$\eta(x,t')$ for $t' > t$.

Two fields $\phi_\eta(x_1,t)$ and $\phi_\eta(x_2,t)$ for the same noises  are  correlated via 
the Langevin equation \eqref{langevin}. 
For a fixed time $\tau$, 
an equal time correlation function 
$\langle \phi_{\eta}(x_1,\tau) \phi_{\eta}(x_2,\tau) \cdots \phi_{\eta}(x_\ell,\tau) \rangle_\eta$
is defined by an average over the noise function: 
\begin{align}
&\langle \phi_{\eta}(x_1,\tau)  \phi_{\eta}(x_2,\tau)\cdots \phi_{\eta}(x_\ell,\tau) \rangle_\eta 
 \notag\\
&\hspace{5mm}= \int [d\eta] \, 
\phi_{\eta}(x_1,\tau)  \phi_{\eta}(x_2,\tau)\cdots \phi_{\eta}(x_\ell,\tau) \, 
\, {\rm exp}\left\{-\frac{1}{4}\int_0^\infty dt \int d^dx\, \eta^2(x,t)\right\},
\label{eq_cor}
\end{align}
where $ [d\eta]$ is normalized so that $\langle 1\rangle_\eta =1$.
As we will see in the next subsection, the correlation function \eqref{eq_cor} can also be written in terms of a $d+1$ dimensional theory with a kind of supersymmetry.

According to Parisi and Wu, the correlation functions of the QFT can be obtained from the large time limit of 
the equal time correlation functions, i.e. 
\begin{align}
\langle  \varphi(x_1)\varphi(x_2) \cdots \varphi(x_\ell) \rangle_{\textrm{QFT}}=
\lim_{\tau\rightarrow \infty} 
\langle \phi_{\eta}(x_1,\tau) \phi_{\eta}(x_2,\tau) \cdots \phi_{\eta}(x_\ell,\tau) \rangle_\eta .
\label{PWtheorem}
\end{align}
Equation \eqref{PWtheorem} is the fundamental relation
in the Parisi-Wu stochastic quantization. 
In the subsection \ref{New proof of Parisi-Wu theorem}, we prove the relation using 
parameter independence of associated supersymmetric theory.

\subsection{Equivalent supersymmetric theory}

The noise average \eqref{eq_cor} associated with the Langevin equation 
is expressed as a normal expectation value of a supersymmetric theory in $d+1$ dimensions using the Nicolai map \cite{NICOLAI1980419}. 
We review how to derive the equivalent $d+1$ dimensional theory in this section. 

Since the solution $\phi_\eta(x,t)$ to the Langevin equation  does not depend on $\eta(x,t')$ for $t' > t$, 
the upper limit of the time integral in Eq.\,\eqref{eq_cor} may be set to $\tau$. 
We then perform a variable change from $\eta$ to $\phi$ using the Nicolai map 
$\eta(x,t)= \partial_t \phi(x,t)  - \square \phi(x,t) + V^\prime(\phi(x,t)) $: 
\begin{align}
&\langle \phi_\eta(x_1,\tau) \cdots \phi_\eta(x_\ell,\tau) \rangle_\eta \nonumber \\
&\hspace{1cm} = \int [d\eta] \, 
\phi_\eta(x_1,\tau) \cdots \phi_\eta(x_\ell,\tau) \, 
\, {\rm exp}\left\{-\frac{1}{4}\int_0^\tau \!\!dt \int d^dx\, \eta^2(x,t)\right\}
\nonumber
\\
&\hspace{1cm} = \int [d\phi] \, 
\phi(x_1,\tau) \cdots \phi(x_\ell,\tau) \, \det(\delta\eta/\delta\phi) \nonumber \\
&\hspace{3cm} \, \times  {\rm exp}\left\{-\frac{1}{4}\int_0^\tau \!\!dt \int d^dx\, ( \partial_t \phi(x,t)  - \square \phi(x,t) + V^\prime(\phi(x,t)))^2 \right\}\, .\label{eq_cor2}
\end{align}
 The Jacobian factor $\det(\delta\eta/\delta\phi)$ can be expressed 
 as a path integral with spinless fermionic variables $\bar\psi$ and $\psi$. 
 
We thus obtain an equivalent description using the $d+1$ dimensional theory with supersymmetry as
\begin{align}
\langle \phi_\eta(x_1,\tau)\cdots \phi_\eta(x_\ell,\tau)  \rangle_\eta =
 \int [d\phi d\bar\psi d\psi] \,
 \phi(x_1,\tau) \cdots \phi(x_\ell,\tau) \, 
 {\rm e}^{-S_0},
\label{cont_step1}
\end{align}
where 
\begin{align}
S_0
 &=  \int_0^\tau \!\!\!dt \int \!d^d x  \, \left\{ \frac{1}{4} \left(\partial_t  
    \phi(x,t) -\square \phi(x,t) +V^\prime(\phi(x,t)) \right)^2 \right. \nonumber \\
 &\hspace{3cm}   \left. + \bar\psi(x,t) \left(\partial_t -\square +V^{\prime\prime}(\phi) \right) 
 \psi(x,t) \frac{}{} \right\}.
\label{susy_action}
\end{align}
We immediately find that the action is invariant under a supersymmetry transformation~\cite{Parisi:1982ud}
\begin{align}
& \delta_\varepsilon \phi(x,t) = \varepsilon\psi(x,t), \label{susy_1}
\\
& \delta_\varepsilon \psi(x,t) = 0,  \\
& \delta_\varepsilon  \bar\psi(x,t) = - \frac{\varepsilon}{2}(\partial_t  
    \phi(x,t) -\square \phi(x,t) +V^\prime(\phi(x,t))), 
\label{susy_3}
\end{align}
where $\varepsilon$ is a Grassmann variable. 


The equivalent theory has asymmetric boundary conditions at two boundaries. 
At $t=0$, $\phi$  satisfies $\phi(x,t)|_{t=0} = \phi_0(x)$ with a fixed initial value $\phi_0(x)$.  
The condition for $\phi$ gives rise to the constraint $\psi(x,t)|_{t=0} =0$. 
$\bar\psi$ does not have any boundary conditions at $t=0$. 
While, at $t=\tau$, $\phi$ is not fixed and is integrated as part of the path integral. 
To understand the boundary condition of $\bar\psi$ at $t=\tau$, we consider a discretized form of fermion kinetic term:
\begin{eqnarray}
\sum_{n=1}^N \bar\psi_n (\psi_n - \psi_{n-1}) =  \bar\psi_N \psi_N - \sum_{n=1}^{N-1} (\bar\psi_{n+1} -\bar\psi_n) \psi_n
\label{free_kin}
\end{eqnarray}
where $n$ denotes the discretized time and the $x$ dependence is dropped for simplicity.
On the RHS of Eq.~\eqref{free_kin}, $\psi_N$ is only included in the first item, and integrating $\psi_N$, we have $\delta(\bar\psi_N)$ on the boundary of $t=\tau$. 
Therefore we may assume $\bar\psi(x,t)|_{t=\tau} =0$, and $\psi$ does not have any boundary conditions at $t=\tau$.
See also appendix \ref{A} for the details of boundary conditions. 
The supersymmetric transformations Eqs.~\eqref{susy_1}-\eqref{susy_3} are consistent with the boundary conditions.

If the boundary condition for time $t$ were periodic, the theory would have another supersymmetry transformation~\cite{Parisi:1982ud}:
\begin{align}
& \delta_{\bar\varepsilon} \phi(x,t) = \bar\varepsilon \bar{\psi}(x,t),
\label{susy2_1}
\\
& \delta_{\bar\varepsilon}  \psi(x,t) = -\frac{\bar \varepsilon}{2} (\partial_t \phi(x,t) +\square \phi(x,t) -V^\prime(\phi(x,t))),   \\
& \delta_{\bar\varepsilon}   \bar\psi(x,t) = 0,  
\label{susy2_3}
\end{align}
where $\bar\varepsilon$ is a Grassmann variable.%
\footnote{This supersymmetric transformation is of the same form as that of $N=2$ 
supersymmetric quantum mechanics~\cite{Witten:1981nf}.  }
The transformation proportional to $\bar\varepsilon$ is, however, broken by
the actual boundary condition obtained from the noise average.

\subsection{Proof of the Parisi-Wu stochastic quantization} 
\label{New proof of Parisi-Wu theorem}

The $d+1$ dimensional theory associated with the Langevin equation was derived in the previous subsection.
The theory is a kind of $N=2$ supersymmetric quantum mechanics for $d=0$ but half of supersymmetry transformations [$\bar\varepsilon\neq 0$ of Eqs.\eqref{susy2_1}-\eqref{susy2_3}]
is broken by the boundary condition. 
In this subsection, we deform the theory changing the boundary condition at $t=0$ 
so that the theory has two supersymmetry invariance exactly. 
We show the Parisi-Wu stochastic quantization through the deformed theory using the time-reversal symmetry.
Then, we show the argument using supersymmetry directly. 
The proofs in this subsection and the treatment of supersymmetry provide a theoretical basis for our formulation in discrete time, 
which will be developed in Section 3.

To define the deformed theory, we first separate the fields defined on the two boundaries from those defined on the bulk.
Let $\varphi_0(x)$ be a boundary field defined only at the boundary of $t=0$ and $\varphi_\tau(x)$ be a boundary field defined only at the boundary of $t=\tau$.
Other $t$-dependent fields are called bulk fields and satisfy the following symmetric boundary conditions:
\begin{align}
\phi(x,t)|_{t=0} = \varphi_0(x), \qquad \psi(x,t)|_{t=0} =0,
\label{bc_0}
\end{align}
and $\bar\psi$ does not have any boundary condition at $t=0$. 
While,   
\begin{align}
\phi(x,t)|_{t=\tau} = \varphi_\tau(x), \qquad \bar\psi(x,t)|_{t=\tau} =0,
\label{bc_tau}
\end{align}
and no boundary condition is assumed for  $\psi$ at $t=\tau$. 

The action of the bulk fields is given by $S_0$ [Eq. \eqref{susy_action}]. 
The boundary fields $\varphi_0$ and $\varphi_\tau$ do not appear explicitly in the bulk action, and are related to bulk fields only 
through the boundary conditions \eqref{bc_0} and \eqref{bc_tau}.
In this setup, a theory that fixes $\varphi_0$ and integrates $\varphi_\tau$ is just the equivalent theory discussed in the previous subsection.

We now change the treatment of the boundary field at $t=0$ from the equivalent theory, 
considering $\varphi_0(x)$ to be a dynamical field
that is integrated by the weight $e^{-S_{\textrm{QFT}}[\varphi_0]}$.\footnote{
A similar idea was proposed in Ref. \cite{Gozzi:1983qxk}.
}
The total action of deformed theory is defined as
\begin{align}
\tilde S \equiv S_{\textrm{QFT}}[\varphi_0]+S_0[\phi,\psi,\bar\psi]\,.
\end{align}
The expectation value of this theory is given by
\begin{align}
\langle {\cal O}  \rangle_{\tilde S} =
 \int [d\varphi_0 d\varphi_\tau][d\phi d\bar\psi d\psi] \,
 \phi(x_1,\tau) \cdots \phi(x_\ell,\tau) \, 
 {\rm e}^{-\tilde S},
\end{align}
where the path integral measure is normalized so that $\langle 1 \rangle_{\tilde S}=1$.  
The bulk fields are integrated to satisfy the boundary conditions given around Eqs.\eqref{bc_0} and \eqref{bc_tau}.

The proof of stochastic quantization using the deformed theory is given as follows. 
The only difference between the equivalent theory of the previous subsection and this variant is the treatment of the boundary condition of $t=0$, 
whose effect on the correlation function of $\varphi_\tau$ vanishes in the large $\tau$ limit: 
\begin{align}
\lim_{\tau\rightarrow \infty}\langle \varphi_\tau(x_1)\cdots \varphi_\tau(x_\ell)  \rangle_{\tilde S} 
=
\lim_{\tau\rightarrow \infty} 
\langle \phi_{\eta}(x_1,\tau) \cdots \phi_{\eta}(x_\ell,\tau) \rangle_\eta\, .
\label{formula_1}
\end{align}
Keeping in mind the time ordering concerning the Langevin equation, we see that 
the bulk field dynamics does not affect the correlation functions of $\varphi_0$.
Since the boundary action of $\varphi_0$ is given by  $S_{\textrm{QFT}}[\varphi_0]$, we have
$\langle \varphi_0(x_1)\cdots \varphi_0(x_\ell)  \rangle_{\tilde S} = \langle \varphi(x_1)\cdots \varphi(x_\ell)  \rangle_{\textrm{QFT}} $. 
Consider the following time reversal, 
$\phi^\prime (t) = \phi(\tau-t), 
\psi^\prime (t) = -\bar\psi(\tau-t), 
\bar\psi^\prime(t) = \psi(\tau-t), 
\varphi^\prime_0 =  \varphi_\tau,
\varphi^\prime_\tau =  \varphi_0
$. 
The boundary conditions are invariant under the time-reversal and the action satisfies 
$ S_{\textrm{QFT}}[\varphi^\prime_0]+S_0[\phi^\prime,\psi^\prime,\bar\psi^\prime] = S_{\textrm{QFT}}[\varphi_0]+S_0[\phi,\psi,\bar\psi] $. 
Therefore, we obtain  
$\langle \varphi_\tau(x_1)\cdots \varphi_\tau(x_\ell)  \rangle_{\tilde S} =
\langle \varphi_0(x_1)\cdots \varphi_0(x_\ell)  \rangle_{\tilde S} $. 
These two results lead to
\begin{align}
\langle \varphi_\tau(x_1)\cdots \varphi_\tau(x_\ell)  \rangle_{\tilde S} =
\langle \varphi(x_1)\cdots \varphi(x_\ell)  \rangle_{\textrm{QFT}}\, .
\label{formula_2}
\end{align}
%
Combining Eq.\eqref{formula_1} and Eq.\eqref{formula_2}, 
the basic formula for the Parisi-Wu stochastic quantization can be derived immediately.

Proofs of stochastic quantization, especially Eq.\eqref{formula_2}, can also be shown explicitly using supersymmetry.
The deformation theory is invariant under two supersymmetry transformations, 
and since it is easier to consider in terms of off-shell supersymmetry, 
we will first redefine the action $\tilde S$ introducing an auxiliary bulk field $H(x,t)$ as follows: 
\begin{align}
\tilde S + \int_0^\tau \!\!dt \int d^d x  \, \left(H +\frac{i}{2}\left(\partial_t  \phi -\square \phi +V^\prime \right) \right)^2 
\rightarrow \tilde S\, .
\end{align}
The explicit form of redefined action $\tilde S$ is given by 
\begin{align}
\tilde S& =  S_{\textrm{QFT}}[\varphi_0]  +  \int_0^\tau \!\!dt \int d^d x  \, \left\{ H^2 + iH\left(\partial_t  \phi -\square \phi +V^\prime \right) 
+ \bar\psi \left(\partial_t -\square +V^{\prime\prime} \right) \psi \right\}.
\label{susy_action_offshell}
\end{align}
For later convenience, we introduce another auxiliary field $\bar H(x,t)$ as a field with the following relation to $H$ as:
\begin{align}
\bar H(x,t) = H(x,t) + i \partial_t \phi(x,t) \, .
\label{Hbar}
\end{align}
These $H$ and $\bar H$ do not satisfy any boundary conditions at two boundaries $t=0$ and $t=\tau$.

Consider two transformations  $Q$ and $\bar Q$ of bulk fields as follows: 


\noindent
\parbox{82mm}{
\vspace{8pt}
\begin{equation}\left\{\,\,
\begin{aligned}
&Q      \phi(x,t)      = \psi(x,t),    \\
&Q      \psi(x,t)      = 0,       \\
&Q      \bar{\psi}(x,t)= -iH(x,t),     \\
&Q      H(x,t)         = 0,       
\end{aligned}\right.
\end{equation}
\vspace{5pt}
}
\parbox{82mm}{
\vspace{8pt}
\begin{equation}\left\{\,\,
\begin{aligned}
&\bar{Q} \phi(x,t)      = \bar{\psi}(x,t), \\
&\bar{Q} \psi(x,t)      = i \bar{H}(x,t), \\
&\bar{Q} \bar{\psi}(x,t)= 0, \\
&\bar{Q} \bar{H}(x,t)   = 0.
\end{aligned}\right.
\end{equation}
\vspace{5pt}
}

\noindent
The transformation law for the boundary fields $\varphi_0(x), \varphi_\tau(x)$ is given 
in a manner consistent with these bulk field transformations and the boundary conditions \eqref{bc_0} and \eqref{bc_tau}.
We therefore have
\begin{align}
Q\varphi_0(x)= \bar Q \varphi_\tau(x) =0.
\label{trivial_transformation}
\end{align}
The transformations satisfy the supersymmetry algebra
\begin{align}
Q^{2}=\bar{Q}^{2} =0, 
\quad \{Q,\bar{Q}\}=-\partial_t\, , 
\label{algebra}
\end{align}
where the relation $\{Q,\bar{Q}\}=-\partial_t$ holds everywhere except at the boundary. 
On the other hand, the nilpotency of $Q$ and $\bar Q$ holds even at the boundary. 
See appendix \ref{A} for more details. 

It is easy to show that  $\tilde S$ satisfies
\begin{align}
\tilde S &= S_{\textrm{QFT}}[\varphi_0] - Q \bar{Q}  \int_0^\tau \!\!dt \, I(t)
\nonumber \\
& = S_{\textrm{QFT}}[\varphi_\tau] + \bar{Q} Q    \int_0^\tau \!\!dt \, I(t),
\label{key_relation}
\end{align}
where
\begin{align}
I(t) = \int  d^d x \left\{ L_{\textrm{QFT}}[\phi(x,t)] + \bar\psi(x,t)\psi(x,t)\right\} \, .
\end{align}
Here  $L_{\textrm{QFT}}[\phi(x,t)]$ is the QFT Lagrangian defined by Eq.\,\eqref{QFT_action}
substituting $\varphi(x) \rightarrow \phi(x,t)$.  
The second equality of Eq.\eqref{key_relation}  is obtained by using the fact $\{Q,\bar{Q}\}=-\partial_t$.
As can be seen from Eq.\,\eqref{key_relation}, 
the deformed theory is invariant under two supersymmetry transformations $Q$ and $\bar{Q}$ with Eq.~\eqref{trivial_transformation}.

To prove Eq.\eqref{formula_2} using supersymmetry, we introduce  
an interpolated action with a parameter $u \in [0,1]$ as \cite{CARDY1985123}
\begin{align}
S_u = \tilde{S} + u\, \bar Q {\cal V}, 
\label{Su}
\end{align}
where
\begin{align}
{\cal V}  =  \int_0^\tau \!\!dt\, \int d^d x \,  \psi(x,t)  \partial_t \phi(x,t) .
\label{V}
\end{align}
We find $S_{u=0}=\tilde{S}$ and $S_{u=1}= S_{\textrm{QFT}}[\varphi_\tau]$ 
up to terms that have no interaction with $\varphi_\tau(x)$ because 
$S_{u=1}$ includes no time derivative terms.\footnote{
This fact is easily understood by writing Eq. \eqref{key_relation} 
as $\tilde S = S_{\textrm{QFT}}[\varphi_\tau] -\bar Q {\cal V} + 
\bar Q Z $ where $Z=\int \!\!dt \, d^d x \{\psi(-i\bar H + L^\prime_{\textrm{QFT}}[\phi])\}$. The $\bar Q$ transformations of 
$\phi,\psi, \bar H$ and $Z$ do not have time derivative terms. The derivative term is only contained in $\bar Q{\cal V}$ which 
is absent in $S_{u=1}$. Since the bulk fields have no propagator in time direction at $u=1$, 
the field dynamics of $\varphi_\tau$ 
is governed only by the boundary action $S_{\textrm{QFT}}[\varphi_\tau]$. 
A careful analysis of the lattice formulation shows that $\bar Q Z$ produce no interaction with $\varphi_\tau(x)$. 
See appendix \ref{A} for more details. 
}
Thus we see that $\tilde S$ and $S_{\textrm{QFT}}[\varphi_\tau]$ are continuously interpolated by $S_u$.

We consider an interpolated expectation value, 
\begin{align}
\displaystyle
C(u)= \frac{\int [d\mu]\, \varphi_\tau(x_1) \varphi_\tau(x_2)  \cdots \varphi_\tau(x_\ell)  
e^{-S_u}} {\int [d\mu]\,  e^{-S_u}}\, ,
\end{align}
where $d\mu \equiv d\varphi_0 d\varphi_\tau d\phi d\psi d\bar\psi d\bar H$ and
the bulk fields satisfy the boundary conditions \eqref{bc_0} and \eqref{bc_tau}.
$C(u)$ interpolates both  sides of Eq.\eqref{formula_2}. So one needs only show $dC/du=0$ to prove Eq.\eqref{formula_2}.
It can be easily shown  
%
\begin{align}
\frac{dC(u)}{du} = -\langle \bar{Q}({\cal O}{\cal V}) \rangle_{S_u} +
\langle {\cal O}\rangle_{S_u} \langle \bar{Q}{\cal V} \rangle_{S_u} \,,  
\label{cont_dcdu}
\end{align}
%
where  ${\cal O}(x_1,x_2,\cdots,x_\ell)  =  \varphi_\tau(x_1)\varphi_\tau(x_2) \cdots \varphi_\tau(x_\ell)$.
Since the action $S_u$ is invariant under $\bar Q$ transformation, we can show $dC/du=0$ using 
 $\int d\mu\,\bar{Q}(\cdots)=0$. 

Therefore, using $\bar Q$ supersymmetry, we can directly show Eq. \eqref{formula_2} indicating 
that the Parisi-Wu stochastic quantization holds.\footnote{
The equality $\langle \varphi_0(x_1)\cdots \varphi_0(x_\ell)  \rangle_{\tilde S} = \langle \varphi(x_1)\cdots \varphi(x_\ell)  \rangle_{\textrm{QFT}} $
can also be proved rigorously using supersymmetry. 
In that case, we can use another interpolated action $S_u = \tilde S+u Q \int_0^\tau \!\!dt\, \int d^d x \,  \bar\psi(x,t)  \partial_t \phi(x,t)$.
Then we can show that $S_{u=0}=\tilde S$ and $S_{u=1}=\tilde S_{\textrm{QFT}}[\varphi_0]$
up to irrelevant terms. }

%
%
\section{Stochastic quantization in discrete time} \label{PWL}
%
%
\subsection{Discrete Langevin equation}
%

We introduce a discrete Langevin time $t=n\epsilon$ for $n \in \mathbb{Z}$ 
with $\epsilon$ being a lattice spacing.  
Let $\phi_n(x) = \phi(x,n\epsilon)$ be a $d+1$ dimensional scalar field  and $\eta_n(x)=\eta(x,n\epsilon)$ be a Gaussian noise function. 
We introduce a backward difference operator $\nabla$ as
\begin{align}
\nabla_{mn} =\frac{1}{\epsilon}\left(\delta_{m,n} - \delta_{m-1,n} \right)
\end{align}
%
for $m,n \in  \mathbb{Z}$, and $\nabla \phi_m \equiv \sum_{n \in \mathbb{Z}} \nabla_{mn} \phi_n$. 
In the following, $m,n$ runs from $1$ to $N$ unless otherwise noted. 

The lattice field $\phi_m$ coincides with $\phi_n $ for fixed $m-n$ in the naive continuum limit $\epsilon \rightarrow 0$, 
so the discretized form of $\delta S/\delta\phi$ (drift force) is not uniquely determined. 
Let $W_n$ be the discretized drift force satisfying $W_n \rightarrow -\square \phi_n(x) +V^\prime(\phi_n(x)).$ in the naive continuum limit $\epsilon \rightarrow 0$. 
%
The Langevin equation \eqref{langevin} is then discretized as
%
\begin{align}
\nabla \phi_n(x) 
 = -W_n(x) +\eta_n(x)
\label{DLeq}
\end{align}
%
for $n=1,2,\cdots, N$, 
with  a fixed initial value $\phi_0$. Eq.~\eqref{DLeq} reproduces the continuum Langevin equation \eqref{langevin} in $\epsilon \rightarrow 0$. 

%
The Parisi-Wu stochastic quantization is obtained by taking the two limits:\footnote{
As discussed in Ref.\cite{batrouni1985langevin}, for finite $\epsilon$, 
the equilibrium action $\bar S$ deviates from the original action $S_{\textrm{QFT}}$ and has a correction term $S_1$ such that
$\bar S=S_{\textrm{QFT}} +\epsilon S_1 + {\it O}(\epsilon^2)$. 
If the discrete Langevin equation consists of local terms that do not break the symmetry of the original theory, the correction term $S_1$ only induces a shift in the parameters of the original action $S$ or a higher order term, and does not affect the physics of the continuous limit 
from the universality hypothesis. 
However, this argument does not hold for nonlocal improvements such as Fourier acceleration or when the fermions are treated dynamically, since nonlocal terms appear in the Langevin equation. With a view to applications in such cases, we discuss here the case of taking the limit of $\epsilon \rightarrow 0$.
}
%
\begin{align}
\langle  \varphi(x_1)\varphi(x_2) \cdots \varphi(x_\ell) \rangle_{\textrm{QFT}}=
\lim_{\tau\rightarrow \infty} \lim_{\epsilon\rightarrow 0}
\langle \phi_{\eta}(x_1,\tau) \phi_{\eta}(x_2,\tau) \cdots \phi_{\eta}(x_\ell,\tau) \rangle_\eta \ ,
\label{PWtheorem_cont_limit}
\end{align}
%
%
where  $\phi_{\eta}$ is a solution of the discrete 
Langevin equation \eqref{DLeq} with $\tau=N\epsilon$. 

Here $\langle \phi_{\eta}(x_1,\tau) \phi_{\eta}(x_2,\tau) \cdots \phi_{\eta}(x_\ell,\tau) \rangle_\eta$ 
has  two scales: $m$ gives the mass scale of the scalar field theory 
and  $\tau$ is another scale introduced for the equal-time correlation function. 
In taking the limit $\epsilon\rightarrow 0$, we need to fix $\tau=N\epsilon$ and $m^2=\tilde m^2 /\epsilon$ 
to constants, where $\tilde m^2$ is a corresponding dimensionless quantity (See Section~\ref{climit} for the details of the continuum limit.)

QFT's correlation functions can be evaluated from numerical simulations with the discrete Langevin equation.
However, performing an extra limit $\epsilon\rightarrow 0$ requires extra numerical costs.

%
\subsection{Weighted noise average and our main theorem}
\label{our_main_result}

The lattice counterpart of Eq.\,\eqref{cont_step1} is obtained 
from the noise average changing variables as $\eta_n(x)=\nabla \phi_n(x)  +W_n(x)$. 
The deformed action is also given on the discrete time, and the first equality of 
Eq.\,\eqref{key_relation} holds.
However, due to the difficulty of lattice supersymmetry, 
it is difficult to satisfy the second equality of Eq.\,\eqref{key_relation}.
%
In supersymmetric quantum mechanics with lattice time (discrete time), 
it is difficult to realize two supersymmetric transformations 
$Q$ and $\bar{Q}$ simultaneously 
\cite{Kato:2013sba,Kadoh:2015zza,Kato:2016fpg,Kato:2018kop,Kadoh:2019bir}.
So this fact implies that the relation \eqref{key_relation} does not hold 
in the discrete time. 


To avoid this difficulty, we consider a modified noise average with a weight function $w$ as
%
\begin{align}
Z_{\eta,w} =  \int [d\eta] \, w[\phi]\, {\rm exp}\bigg\{-\frac{\epsilon}{4} \sum_{n=1}^N  \int d^dx\, \eta_n^2(x)\bigg\}, 
\label{reweighted_Z}
\end{align}
and the expectation value is 
\begin{align}
\langle\, {\cal O}[\phi] \,\rangle_{\eta, w} = \frac{1}{Z_{\eta,w}} 
 \int [d\eta] \, {\cal O}[\phi]\,w[\phi]\,  {\rm exp}\bigg\{-\frac{\epsilon}{4} \sum_{n=1}^N  \int d^dx\, \eta_n^2(x)\bigg\},  
\label{reweighted_O}
\end{align}
%
where $\phi$ is a solution of discrete Langevin equation \eqref{DLeq}. 

The weight function is defined from two different discretizations of drift forces, $W_n$ and $\overline W_n$. 
$W_n(x)$ is a lattice drift force giving the discrete Langevin equation \eqref{DLeq}
and $\overline W_n(x)$ is another discrete expression of $\delta S/\delta \phi$ used to define the weight function. 
Both  $W_n(x)$ and $\overline W_n(x)$ are assumed to depend only on $\phi_n(x), \phi_{n-1}(x)$: 
\begin{align}
W_n(x) = W(\phi_n(x), \phi_{n-1}(x)), \nonumber \\
\overline W_n(x) = \overline W(\phi_{n-1}(x), \phi_{n}(x)) , 
\label{w_constraint_1}
\end{align} 
with the desirable continuum limits,
\begin{align}
W_n, \overline W_n \rightarrow -\square \phi_n(x) +V^\prime(\phi_n(x)).
\label{w_constraint_2}
\end{align} 
Suppose further that the following theoretical constraints are imposed on $W_n, \overline W_n$: 
\begin{align} 
{\rm det} M, \,  {\rm det} \overline M >0, \qquad \frac{{\rm det} \overline M}{{\rm det} M}=1 + {\it O}(\epsilon), 
\label{w_constraint_3}
\end{align} 
where 
\begin{align}
& M(x,m;y,n)= \left\{ \nabla_{mn} +\frac{\partial W_m(x)}{\partial \phi_n(x)} \right\} \delta^d(x-y), 
\label{M_def}
\\
& \overline M(x,m;y,n) = \left\{  {\nabla}^T_{mn} +  \frac{\partial {\overline W}_{m}(x)}{\partial \phi_{n-1}(x)} \right\}
\delta^d(x-y). 
\label{barM_def}
\end{align}
Note that $\nabla^T_{mn}=\nabla_{nm}=- (\delta_{m+1,n} - \delta_{m,n})/\epsilon$ is the inverse sign of forward difference operator. 
In Section \ref{toy}, we present some solutions that satisfy these constraints.

Besides these assumptions, the assumption such as the existence of a gap between the ground state of the Fokker-Planck Hamiltonian and the energy eigenvalue of the first excited state, which is required for the proof using the Fokker-Planck equation, is implicitly imposed here as well.

Our main theorem in this paper is given by
%
\begin{align}
\langle  \varphi(x_1)\varphi(x_2) \cdots \varphi(x_\ell) \rangle_{\textrm{QFT}}=
\lim_{N \rightarrow \infty} \langle \phi_{\eta,N}(x_1) \phi_{\eta,N}(x_2) 
\cdots \phi_{\eta,N}(x_\ell) \rangle_{\eta,w}\,,\label{latticePW}
\end{align}
%
where the weighted noise average $ \langle  \cdots \rangle_{\eta,w}$ is defined by 
Eq.\,\eqref{reweighted_O} with 
%
\begin{align}
w[\phi] &= \frac{ {\rm det} \overline M }{  {\rm det} M } \ 
{\rm exp} \bigg\{\frac{\epsilon}{4} \sum_{n=1}^N \int d^dx
\left[
    W_n^2(x)-\overline{W}_{n}^2(x)
    +2 \nabla \phi_n(x) \big(W_n(x)+\overline{W}_{n}(x)\big)
\right]
   \nonumber \\
&  
\hspace{2cm}
\frac{}{}
- S_{\textrm{QFT}}[\phi_N]+ S_{\textrm{QFT}} [\phi_0]
\bigg\}.
\label{weight}
\end{align}
The $\bar Q$ symmetry was important to show Eq. \eqref{formula_2} in the continuum time.  
As will be seen in the next subsection, the weight function $w$ is chosen so that the deformed theory 
can be written as $\tilde S_{ \textrm{LAT} }=S_{\textrm{QFT}}[\phi_N]+\bar Q \bar X$ in discrete time
with an exact $\bar Q$ symmetry.

The condition on $W$ in Eq. \eqref{w_constraint_1} guarantees that the discrete Langevin equation \eqref{DLeq} coincides with Eq. \eqref{langevin} in the naive continuum limit $\epsilon \rightarrow 0$. 
Because of ${\rm det} \overline{M}>0$ in \eqref{w_constraint_3}, $w$ will never be zero and the expectation value \eqref{reweighted_O} is well-defined. The conditions \eqref{w_constraint_1} and ${{\rm det} \overline M}/{{\rm det} M}=1 + {\it O}(\epsilon)$
in Eq. \eqref{w_constraint_3} guarantee that $w=1$ in $\epsilon \rightarrow 0$.
The condition on $\overline{W}$ in Eq. \eqref{w_constraint_1} and ${\rm det} M>0$  in Eq.~\eqref{w_constraint_3}
will be used in the proof of the following subsection.

Compare Eq.~\eqref{latticePW} with Eq.~\eqref{PWtheorem_cont_limit}. 
The relation \eqref{latticePW} does not have the continuum limit of $\epsilon \to 0$.
This is an economic merit of our lattice formulation to reduce numerical costs.
Since $w=1$ in $\epsilon \rightarrow 0$, the weighted method 
is meaningful only in the discrete-time formulation.

%
\subsection{Proof of our main theorem}
%

The Parisi-Wu stochastic quantization holds after taking the continuum limit  $\epsilon \to 0$
for the discrete Langevin time.
In this section, it is shown that the similar statement \eqref{latticePW} also holds 
\textit{without} taking the continuum limit.

We first derive a $\bar Q$ invariant theory ($\tilde S_{ \textrm{LAT} }=S_{\textrm{QFT}}[\phi_N]+\bar Q \bar X$) 
from the weighted noise average,
after which  we will prove Eq.\eqref{latticePW} using the $\bar Q$ symmetry following the same strategy as 
in the proof for continuum time given in Subsection \ref{New proof of Parisi-Wu theorem}. 

Performing calculations similar to Eq.~\eqref{eq_cor2} leads to 
\begin{align}
&\langle \phi_{\eta,N}(x_1) \cdots \phi_{\eta,N}(x_\ell) \rangle_{\eta,w} 
\nonumber \\
&\hspace{1cm} = \frac{1}{Z_{\eta,w}} \int [d\eta] \, 
\phi_{\eta,N}(x_1) \cdots \phi_{\eta,N}(x_\ell) \, w[\phi_\eta] \,
\, {\rm exp}\left\{-\frac{\epsilon}{4}\sum_{n=1}^N  \int d^dx\, \eta^2_n(x)\right\}
\nonumber
\\
&\hspace{1cm} = \frac{1}{Z_{\eta,w}} \int [d\phi] \, 
\phi_N(x_1) \cdots \phi_N(x_\ell) \, {\rm det}M\, w[\phi] \, 
 {\rm exp}\left\{-\frac{\epsilon}{4} \sum_{n=1}^N  \int d^dx\, ( \nabla \phi_n(x)  + W_n(x) )^2 \right\}
\nonumber \\
&\hspace{1cm} =\frac{1}{Z_{\eta,w}} 
\int [d\phi d\bar\psi d\psi] \,
 \phi_N(x_1) \cdots \phi_N(x_\ell) \,  {\rm e}^{-S_{\textrm{LAT}}},
\label{lat_step1}
\end{align}
where 
%
\begin{align}
S_{\textrm{LAT}} 
&=  \epsilon \sum_{n=1}^{N} \int d^d x  \, \frac{1}{4} \left( \nabla \phi_n(x) 
    -\overline {W}_n(x) \right)^2
    -  \epsilon \sum_{m,n=1}^{N} \int d^d x \, \psi_m(x) \overline{M}_{mn}(x) \bar\psi_n(x)  \nonumber \\
 & \hspace{7mm}
   + S_{\textrm{QFT}}[\phi_N] - S_{\textrm{QFT}} [\phi_0] \ ,
\label{lat_action}
\end{align}
with $\overline{M}_{mn}(x)= {\nabla}^T_{mn} +  \frac{\partial {\overline W}_{m}(x)}{\partial \phi_{n-1}(x)}$ and 
%
%
$
[d\phi d\bar\psi d\psi] = \prod_{x \in \mathbb{R}^d} 
\prod_{n=1}^N d\phi_n(x) d\bar\psi_n(x) d\psi_n(x) 
 $.

In Eq. \eqref{lat_step1}, the condition on $\overline{W}$ in Eq. \eqref{w_constraint_1}
and  ${\rm det} M>0$  in Eq.~\eqref{w_constraint_3} are used. 
We emphasize again that $Z_{\eta,w}$ is non-zero because of ${\rm det} \overline{M}>0$ which is assumed in Eq.~\eqref{w_constraint_3}, 
and the expectation value is well-defined.

%

%
We now define a deformed theory $\tilde S_{ \textrm{LAT} }$, which is similar to 
Eq.\,\eqref{susy_action_offshell}, 
treating $\phi_0$ as a dynamical field integrated with the weight $e^{-S[\phi_0]}$: 
%
\begin{align}
\tilde S_{ \textrm{LAT} } \equiv S_{  \textrm{LAT} } + S_{\textrm{QFT}}[\phi_0]\, .
\end{align}
%
The expectation value of the theory  $\tilde S_{ \textrm{LAT} }$ is given by the path integral with ${\rm exp}(-\tilde S_{ \textrm{LAT} })$ 
and is denoted as $\langle \cdots \rangle_{\tilde S_{ \textrm{LAT} }}$.
As with Eq.~\eqref{formula_1} in the continuum theory, the only difference between $S_{ \textrm{LAT} }$ and $\tilde S_{ \textrm{LAT} }$
is the treatment of the boundary field at $t = 0$, 
so the following equality holds:
\begin{align}
\lim_{N\rightarrow \infty} 
\langle \phi_N(x_1) \cdots \phi_N(x_\ell) \rangle_{\tilde S_{ \textrm{LAT} }}
=
\lim_{N\rightarrow \infty}\langle \phi_{\eta,N}(x_1)\cdots \phi_{\eta,N}(x_\ell)  \rangle_{\eta,w} 
\label{lat_formula_1}
\end{align}
for a fixed $\epsilon$.

To make the off-shell $\bar Q$ symmetry clear, we now redefine $\tilde S_{ \textrm{LAT} }$ 
by introducing an auxiliary field $\bar H_n$ as follows. 
\begin{align}
\tilde S_{ \textrm{LAT} } + \epsilon\sum_{n=1}^{N} \int d^d x \left(\bar H_n(x) - \frac{i}{2}(\nabla \phi_n(x) - \overline W_n(x)) \right)^2  \rightarrow \tilde S_{  \textrm{LAT} }\, .
\end{align}
The explicit expression of $\tilde S_{ \textrm{LAT}}$ (with the auxiliary field $\bar H$) is 
\begin{align}
\tilde S_{  \textrm{LAT} } & =  S_{\textrm{QFT}}[\phi_N] + \epsilon \sum_{n=1}^{N} \int d^d x
\left\{ \bar H_n^2(x) - i \bar H_n(x) (\nabla \phi_n(x) - \overline W_n(x) ) \right\}  \nonumber \\
& - \epsilon \sum_{m,n=1}^{N} \int d^d x\, \psi_m(x) 
{\overline M}_{mn}(x) \bar\psi_n(x)  \, .
\end{align}
The expectation value in the deformed theory 
is given by 
%
\begin{align}
\langle \phi_{N}(x_{1})\phi_{N}(x_{2}) \cdots \phi_{N}(x_\ell) \rangle_{\tilde{S}_{  \textrm{LAT} }}
 = \frac{ \int d\tilde \mu\,
          \phi_{N}(x_{1})\phi_{N}(x_{2}) \cdots \phi_{N}(x_\ell)\,
           e^{-\tilde{S}_{  \textrm{LAT} }} }{ \int d\tilde \mu \,
           e^{-\tilde{S}_{  \textrm{LAT} }} }\,,
\end{align}
%
where $d\tilde \mu =  \prod_{x \in \mathbb{R}^d} d\phi_0(x) \prod_{n=1}^{N}   d\phi_n(x)d\bar\psi_n(x) d\psi_n(x)  d\bar H_n(x) $. 

Consider the following 
$\bar Q$ transformations which satisfy  $\bar Q^2=0$: 
\begin{align}
&\bar Q \phi_{n-1}(x) = \bar \psi_n(x),\\
&\bar Q \psi_n(x) = i\bar H_n(x) , \\ 
&\bar Q \bar \psi_n(x) = 0,  \\
&\bar Q \bar H_n(x) =0
\end{align}
for $n=1,2,\cdots,N$,
and $\bar Q \phi_N=0$.
We immediately find that $\tilde S_{  \textrm{LAT} }$ is $\bar Q$-invariant because 
the bulk action can be expressed in a $\bar Q$-exact form as
\begin{align}
\tilde S_{  \textrm{LAT} } &= S_{\textrm{QFT}}[\phi_N] + \bar Q  \left\{\epsilon \sum_{n=1}^{N} \int d^d x  \, \big[
-\nabla \phi_n(x) -i\bar H_n(x) +\overline{W}_n(x) \big]\psi_n(x) \right\} ,
 \label{lat_susy_action_Qexact}
\end{align}
and $S_{\textrm{QFT}}[\phi_N]$ is $\bar Q$-closed.

Since Eq.~\eqref{lat_formula_1}, which is a lattice version of Eq.~\eqref{formula_1}, holds, 
we know that Eq.~\eqref{latticePW} can be shown immediately if the following relation holds.
\begin{align}
\langle \phi_N(x_1)\cdots \phi_N(x_\ell)  \rangle_{\tilde S_{ \textrm{LAT}}} =
\langle \varphi(x_1)\cdots \varphi(x_\ell)  \rangle_{\textrm{QFT}}
\label{lat_formula_2}
\end{align}
which corresponds to Eq.~\eqref{formula_2}. To show this, we consider $\bar Q$ symmetry of the deformed theory.

We introduce $u \in [0,1]$ and define 
an interpolated action
%
\begin{align}
S_u = \tilde S_{  \textrm{LAT} } + u\, \bar Q\, {\cal V}  
\label{lattieSu}
\end{align}
%
where 
\begin{align}
&{\cal V}  =  \epsilon \sum_{n=1}^{N} \int d^d x
\big[ \nabla \phi_n(x)-\overline{W}_n(x)+L_{\textrm{QFT}}^\prime(\phi_{n-1}(x))\big]
\psi_{n}(x),
\label{lattieV}
\\
&L_{\textrm{QFT}}^\prime(\phi_n(x)) = -\square \phi_{n}(x) + V'(\phi_{n}(x)).
\end{align}
The second and third terms of Eq.\,\eqref{lattieV} will cancel each other
in the continuum limit, and $S_{u}$ in Eq.\,\eqref{lattieSu} reduces to
Eq.\,\eqref{Su}.
The associated expectation value is
%
\begin{align}
C(u) \equiv \frac{ \int  d\tilde \mu  \, 
{\cal O}(x_1,x_2,\cdots, x_\ell) 
\, {\rm e}^{-S_u}}
{ \int  d\tilde \mu \,   {\rm e}^{-S_u}}\, , 
\end{align}
%
where  ${\cal O}(x_1,x_2,\cdots, x_\ell)  =  \phi_N(x_1)\phi_N(x_2) \cdots \phi_N(x_\ell)$.

From the definition, it can be seen that $C(u)$ interpolates two sides of Eq.~\eqref{lat_formula_2} because
$S_{u=0}=\tilde S_{  \textrm{LAT} }$ and
\begin{align}
S_{u=1} = S_{\textrm{QFT}}[\phi_N] + \sum_{n=1}^{N} \int d^d x
\left( \frac{1}{4} L^\prime (\phi_{n-1}(x))^2 -\psi_{n}(x) L^{\prime\prime}(\phi_{n-1}(x))\bar \psi_{n}(x) 
\right)
\end{align}
after integrating $\bar H_n$. 
We should notice that $\phi_N$ is only contained in the first term. 
The limit of $u \rightarrow 1$ can be safely taken because  
the second term provides well-defined damping terms for the bulk fields $\phi_n (n=0,1,\cdots,N-1)$. 

The remaining task here is to show $dC/du=0$ to prove Eq.~\eqref{lat_formula_2}. 
We can show that
%
\begin{align}
\frac{dC(u)}{du} = - \langle \bar Q ({\cal O V})  \rangle_{S_u} +
\langle{\cal O} \rangle_{S_u}  \langle \bar Q {\cal V} \rangle_{S_u} =0,
\end{align}
where $\langle\cdots \rangle_{S_u}\!$ is the expectation value with the action $S_u$.
The last equality comes from $\int\! d\mu \bar Q(\cdots) \!=\!0$. 
We thus find that Eq.~\eqref{lat_formula_2} is correct and our main theorem holds.

%

\section{ Tests in a toy model in $d=0$}\label{toy}

\subsection{Model}

We demonstrate how our method works with a zero-dimensional toy model for $x \in \mathbb{R}$,  
\begin{equation}
Z=\int_{-\infty}^\infty \,dx\, {e}^{-V(x)}.
\end{equation}
To be concrete we take
\begin{equation}
V(x)=\frac{1}{2}m^2\, x^2 + \frac{1}{4}\lambda\, x^4,
\label{0dmodel}
\end{equation}
where $m^2$ and $\lambda$ are the model parameters that imitate mass$^2$ and coupling constant, respectively. 
The observables are expectation values of $x^{2p}$ with positive integer $p$
\begin{equation}
\langle x^{2p} \rangle_x = \frac{1}{Z}\int_{-\infty}^\infty \, dx\, x^{2p} {e}^{-V(x)},
\end{equation}
which gives the left-hand side of Eq.~\eqref{latticePW}.  
We will examine if $\langle x^{2p} \rangle_x $ is reproduced from the large $N$ limit 
of $\langle \phi_N{}^{2p}\rangle_{\eta,w}$ using the perturbation theory (Subsection \ref{PT_test}) and numerical simulations 
(Subsection \ref{num_test}).

The discretized Langevin equation (\ref{DLeq}) is
\begin{equation}
\frac{\phi_n-\phi_{n-1}}{\epsilon} = -W_n +\eta_n,\label{0dDL}
\end{equation}
for $n=1,2,\cdots,N$. The  lattice white noise is normalized as 
\begin{equation}
\langle\eta_n \eta_{n'}\rangle_{\eta}=\frac{2}{\epsilon}\,\delta_{n,n'}.
\end{equation}
The choice of the drift force $W_n$ can be arbitrary as far as it satisfies three properties explained in Section~3. 
We will consider two cases:

\begin{itemize}

\item A-type
\begin{equation}
W_n^{\rm A} \equiv \frac{m^2}{2}(\phi_n+\phi_{n-1})+\frac{\lambda}{2}(\phi_n^3+\phi_{n-1}^3).
\label{V'_ST}
\end{equation}

\item B-type
\begin{equation}
W_n^{\rm B} 
\equiv \frac{m^2}{2}(\phi_n+\phi_{n-1})+\frac{\lambda}{4}(\phi_n^3+\phi_n^2\phi_{n-1}+\phi_n\phi_{n-1}^2+\phi_{n-1}^3).
\label{V'_CLR}
\end{equation}

\end{itemize}
It can be easily shown that both drift forces satisfy three conditions \eqref{w_constraint_1}, \eqref{w_constraint_2} and \eqref{w_constraint_3} in Section~3.

We take $\overline W_n$ needed to define the weight factor \eqref{weight} as $\overline W_n^{i} = W_n^i$ for $i=$ A, B.
In the case, we have 
\begin{equation}
w^{\rm A}=\frac{1+\epsilon\left(\frac{m^2}{2}+\frac{3}{2}\lambda\phi_0^2\right)}{1+\epsilon\left(\frac{m^2}{2}+\frac{3}{2}\lambda\phi_N^2\right)}
\exp\left\{\frac{\lambda}{2}\left(\sum_{n=1}^N\, (\phi_n-\phi_{n-1}) (\phi_n{}^3+\phi_{n-1}{}^3)-\frac{\phi_N{}^4}{2}+\frac{\phi_0{}^4}{2}\right)\right\},
\label{w_ST}
\end{equation}
and 
\begin{equation}
w^{\rm B}=\prod_{n=1}^N \frac{ 1+\epsilon\left(\frac{m^2}{2}+\frac{\lambda}{4}(\phi_n{}^2+2\phi_n\phi_{n-1}+3\phi_{n-1}{}^2)\right)}{
1+\epsilon\left(\frac{m^2}{2}+\frac{\lambda}{4}(3\phi_n{}^2+2\phi_n\phi_{n-1}+\phi_{n-1}{}^2)\right)}.
\label{w_CLR}
\end{equation}
Note that $w^{\rm A}$ has one fraction coming from the determinant part and the sum of $n=1,2,\cdots, N$ in the exponential part. In contrast, $w^{\rm B}$ is given by the product of $N$ fractions in the determinant part and has no exponential function.

Finally, we make some miscellaneous comments on the two types of drift forces examined in this paper. 
The A-type drift term is given by the average of the endpoint values, as follows
\begin{equation}
W_n^{\rm A} = \frac{V'(\phi_n)+V'(\phi_{n-1})}{2}.
\end{equation}
This is inspired by the Stratonovich integral of the stochastic process, though we will not use continuum stochastic calculus.
On the other hand, the B-type drift term can be given as follows, 
\begin{equation}
W_n^{\rm B} = \frac{V(\phi_n)-V(\phi_{n-1})}{\phi_n-\phi_{n-1}}.
\end{equation}
We find that $W_n^{\rm B}$ satisfies
\begin{equation}
\epsilon\sum_{n=1}^N \nabla\phi_n\,W_n^{\rm B}=V(\phi_N)-V(\phi_0),
\end{equation}
which is   a corollary of the Cyclic Leibniz Rule (CLR)~\cite{Kato:2013sba, Kadoh:2015zza, Kato:2016fpg, Kato:2018kop, Kadoh:2019bir} 
and is the reason why   $w^{\rm B}$ has no exponential function.

\subsection{Continuum limit}\label{climit}
Before going into an actual analysis of the model, here we will discuss the continuum limit. 
To do so, we must distinguish dimensionful variables and parameters from dimensionless ones.
Note that, regardless of the value of $d$, the fictitious time $t$ has its own dimension. Let us denote it as $[t]\sim {\sf T}$ with dimensional symbol ${\sf T}$. The lattice constant $\epsilon$ for discretized fictitious time also has $[\epsilon]\sim {\sf T}$. Then for the Gaussian white noise $\eta$ in $d=0$ case we have $[\eta]\sim {\sf T}^{-1/2}$ because weight factor is $\exp(-\frac{1}{4}\int dt\, \eta(t)^2)$ or, in the discrete form, $\exp(-\frac{1}{4}\epsilon\sum_n \eta_n{}^2)$.
For the other variables and parameters, from the simplest Langevein equation, 
\begin{equation}
\frac{1}{\epsilon}(\phi_n-\phi_{n-1})=-m^2\phi_{n-1}-\lambda\phi_{n-1}^3+\eta_n, 
\label{naiveL}
\end{equation}
we can easily see $[\phi]\sim {\sf T}^{1/2}$, $[m^2]\sim {\sf T}^{-1}$ and $[\lambda]\sim {\sf T}^{-2}$.

Let us introduce dimensionless variables and parameters on the lattice by putting a tilde on each symbol:
\begin{equation}
\tilde\eta_n=\sqrt{\epsilon}\,\eta_n,\quad \tilde\phi_n=\frac{1}{\sqrt{\epsilon}}\phi_n,\quad
\tilde m^2=\epsilon\, m^2,\quad\tilde\lambda=\epsilon^2\lambda.
\label{dimensionless_quantities}
\end{equation}
Then Eq.(\ref{naiveL}) can be rewritten in a fully dimensionless way:
\begin{equation}
\tilde\phi_n-\tilde\phi_{n-1} = -\tilde m^2\tilde\phi_{n-1}-\tilde\lambda\tilde\phi_{n-1}^3 +\tilde\eta_n.
\end{equation}
Similarly, all equations and weights with the A-type and B-type drift terms can be expressed only by the dimensionless quantities \eqref{dimensionless_quantities}.
The actual lattice calculation uses this type of dimensionless equations.

Our observable is, for example, a $2p$-point function
\begin{equation}
\langle (\phi_N)^{2p}\rangle_{\eta}=\epsilon^p\langle (\tilde\phi_N)^{2p}\rangle_{\tilde\eta}
\end{equation}
that is going to coincide with $\langle x^{2p}\rangle_x$ in the $N\rightarrow\infty$ limit.
The quantity $\langle (\tilde\phi_N)^{2p}\rangle_{\tilde\eta}$ is a function of $\tilde m^2$ and $\tilde\lambda$, or alternatively $\tilde m^2$ and $\frac{\tilde\lambda}{\tilde m^4}$ ($=\frac{\lambda}{m^4}$).
So we can write
\begin{equation}
m^{2p}\langle (\phi_N)^{2p}\rangle_{\eta}=\tilde m^{2p}\langle (\tilde\phi_N)^{2p}\rangle_{\tilde\eta}=f(\frac{\tilde\lambda}{\tilde m^4},\tilde m^2)=f(\frac{\lambda}{m^4},\epsilon\,m^2)\label{dependence}
\end{equation}
with some function $f$. Therefore, the continuum limit $\epsilon\rightarrow 0$ is to take the limit $\tilde m^2\rightarrow 0$ while $\frac{\tilde\lambda}{\tilde m^4}$ is fixed in the dimensionless lattice quantity: 
\begin{equation}
\lim_{\underset{\lambda,m^2,\tau \textrm{ fixed}}{\epsilon\rightarrow 0}}m^{2p}\langle (\phi_N)^{2p}\rangle_{\eta}
=\lim_{\underset{\frac{\tilde\lambda}{\tilde m^4},\tau \textrm{ fixed}}{\tilde m^2\rightarrow 0}}\tilde m^{2p}\langle (\tilde\phi_N)^{2p}\rangle_{\tilde\eta}.
\end{equation}
where $\tau = N\epsilon$. 
Eq.~\eqref{PWtheorem_cont_limit} means that the large $\tau$ limit of the above equation coincides with $m^{2p} \langle x^{2p} \rangle_x$ for the same fixed value of $\lambda/m^4$.

On the other hand, our method explained in Section 3 gives us a simpler way to relate continuum quantity and lattice one. Namely a function in Eq.(\ref{dependence}) only depends on $\frac{\tilde\lambda}{\tilde m^4}$,
\begin{equation}
m^{2p}\langle (\phi_N)^{2p}\rangle_{\eta, w}=\tilde m^{2p}\langle (\tilde\phi_N)^{2p}\rangle_{\tilde\eta, w}=f(\frac{\tilde\lambda}{\tilde m^4})=f(\frac{\lambda}{m^4}).
\end{equation}
That means we do not need to take the $\epsilon\rightarrow 0$ limit explicitly. 
It is the claim of our theorem \eqref{latticePW} that the large $N$ limit of the above equation reproduces $m^{2p} \langle x^{2p} \rangle_x$ for fixed $\lambda/m^4$.
In this case, $\epsilon$ only appears in the conversion relation to the dimensionful quantity $\frac{1}{\epsilon}\tilde m^2=m^2$ and can be fixed to any value.

\subsection{Perturbation theory}
\label{PT_test}
  \newcommand{\tphi}{\tilde\phi}
  \newcommand{\teta}{\tilde\eta}
  \newcommand{\tms}{\tilde m^2}
  \newcommand{\tlmd}{\tilde\lambda}

Let us make a perturbative analysis in this subsection.
The exact expression of the perturbative $2p$-point function is
\begin{equation}
\langle x^{2p}\rangle_x = \frac{(2p-1)!!}{m^{2p}}\left[1-\frac{\lambda}{m^4}p(p+2)+O(\lambda^2)\right],
\label{2pexact}
\end{equation}
which we want to derive.

We take the B-type for the explanation. Defining $b=\frac{2-\tms}{2+\tms}$ and $c=\frac{2}{2+\tms}$, 
the B-type discrete Langevin equation (\ref{0dDL}) with Eq.(\ref{V'_CLR}) can be put into the form
\begin{equation}
\tphi_n = b \tphi_{n-1}+c \teta_n
-\tlmd \frac{c}{4}(\tphi_n{}^3+\tphi_n{}^2\tphi_{n-1}+\tphi_n\tphi_{n-1}{}^2+\tphi_{n-1}{}^3).\label{0dDL1}
\end{equation}
We will solve this in a series of $\tlmd$ expansions:
\begin{equation}
\tphi_n=\tphi_n^{(0)}+\tlmd\tphi_n^{(1)}+\tlmd^2\tphi_n^{(2)}+\cdots.
\end{equation}
Plugging this into the above equation we obtain the following relations in each order of $\tlmd$,
\begin{eqnarray}
O(\tlmd^0):\quad\tphi_n^{(0)}&=&b\tphi_{n-1}^{(0)}+c\teta_n,\label{0theq}\\
O(\tlmd^1):\quad\tphi_n^{(1)}&=&b\tphi_{n-1}^{(1)}-\frac{c}{4}(\tphi_n^{(0)}{}^3+\tphi_n^{(0)}{}^2\tphi_{n-1}^{(0)}+\tphi_n^{(0)}\tphi_{n-1}^{(0)}{}^2+\tphi_{n-1}^{(0)}{}^3),\label{1steq}\\
&\vdots&\nonumber
\end{eqnarray}
By solving these equations we can determine $\tphi_n^{(k)}$  order by order the details of which will be given in Appendix~\ref{A.perturb}.

Using these variables we can evaluate the $2p$-point function without weights
\begin{equation}
\langle\tphi_N{}^{2p}\rangle_{\teta}
=\langle\tphi_N^{(0)}{}^{2p}\rangle_{\teta}
+2p\tlmd\langle\tphi_N^{(0)}{}^{2p-1}\tphi_N^{(1)}\rangle_{\teta}
+O(\tlmd^2).
\end{equation}
Taking a limit $N\rightarrow\infty$ we eventually obtain
\begin{equation}
\tms{}^p\langle\tphi_N{}^{2p}\rangle_{\teta}^{\rm B} \underset{N\rightarrow\infty}{\longrightarrow}(2p-1)!!\left(1-\frac{\tlmd}{\tilde m^4}\frac{4p+p^2(2+\tms)}{2+\tms}+O(\tlmd^2)\right).\label{2pCLR}
\end{equation}
The right-hand side depends not only $\frac{\tlmd}{\tilde m^4}$ but also $\tms$, so we need the continuum limit $\tms\rightarrow 0$ with $\frac{\tlmd}{\tilde m^4}$ fixed, as explained in the previous subsection, to reproduce the perturbative result of the zero dimensional integral (\ref{2pexact}).

Now we turn to include the weight factor.
The weight factor $w$ for the B-type can  be expanded concerning the $\tlmd$ as
\begin{equation}
w^{\rm B} = 1-\frac{\tlmd}{2+\tms}\tphi_N^{(0)}{}^2 +O(\tlmd^2).
\end{equation}
Making use of this, we have 
\begin{eqnarray}
\langle\tphi_N{}^{2p}\rangle^{\rm B}_{\teta,w}&=&\frac{\langle\tphi_N{}^{2p}w\rangle_{\teta}}{\langle w\rangle_{\teta}} \nonumber\\
&=&\langle\tphi_N^{(0)}{}^{2p}\rangle_{\teta}
+\tlmd\left[2p\langle\tphi_N^{(0)}{}^{2p-1}\tphi_N^{(1)}\rangle_{\teta}
-\frac{1}{2+\tms}\left(\langle\tphi_N^{(0)}{}^{2p+2}\rangle_{\teta}-\langle\tphi_N^{(0)}{}^{2p}\rangle_{\teta}\langle\tphi_N^{(0)}{}^2\rangle_{\teta}
\right)\right]\nonumber\\
& & \hspace{30mm}+O(\tlmd^2).
\end{eqnarray}
The correction term coming from the weight factor is evaluated as
\begin{equation}
-\frac{\tlmd}{2+\tms}\left(\langle\tphi_N^{(0)}{}^{2p+2}\rangle_{\teta}-\langle\tphi_N^{(0)}{}^{2p}\rangle_{\teta}\langle\tphi_N^{(0)}{}^2\rangle_{\teta}
\right)
\underset{N\rightarrow\infty}{\longrightarrow}-\frac{\tlmd}{2+\tms}\frac{2p(2p-1)!!}{\tilde m^{2p+2}}.
\end{equation}
Combining Eq.(\ref{2pCLR}), this leads us to the result
\begin{equation}
\langle\tphi_N{}^{2p}\rangle_{\teta,w}^{\rm B} \underset{N\rightarrow\infty}{\longrightarrow}\frac{(2p-1)!!}{\tilde m^{2p}}\left(1-\frac{\tlmd}{\tilde m^4}p(p+2)+O(\tlmd^2)\right).
\end{equation}
Alternatively, for the dimensionful quantities, we have
\begin{equation}
\langle\phi_N{}^{2p}\rangle_{\eta,w}^{\rm B} \underset{N\rightarrow\infty}{\longrightarrow}\frac{(2p-1)!!}{m^{2p}}\left(1-\frac{\lambda}{m^4}p(p+2)+O(\lambda^2)\right),
\end{equation}
which exactly coincides with the desired result (\ref{2pexact}) up to the first order of $\lambda$ without taking any continuum limit.

The same analysis can be extended to the A-type. 
Weight factor, in this case, can be expanded in terms of $\tlmd$ as
\begin{equation}
w^{\rm A} = 1 + \tlmd \left[ 
\frac{1}{2}\sum_{n=1}^N(\tphi_n^{(0)}-\tphi_{n-1}^{(0)})(\tphi_n^{(0)}{}^3+\tphi_{n-1}^{(0)}{}^3)-\frac{1}{4}\tphi_N^{(0)}{}^4-\frac{3}{2+\tms}\tphi_N^{(0)}{}^2 
\right]+ O(\tlmd^2).
\end{equation}
In general, if the weight factor has an expansion in terms of the coupling constant $\tlmd$ like
\begin{equation}
w=1+\tlmd\, w^{(1)} +O(\tlmd^2),
\end{equation}
then $2p$-point function can be evaluated up to $O(\tlmd)$ according to the following expansion
\begin{eqnarray}
\langle\tphi_N{}^{2p}\rangle_{\teta,w}&=&\frac{\langle\tphi_N{}^{2p}w\rangle_{\teta}}{\langle w\rangle_{\teta}} \nonumber\\
&=&\langle\tphi_N^{(0)}{}^{2p}\rangle_{\teta}
+\tlmd\left[2p\langle\tphi_N^{(0)}{}^{2p-1}\tphi_N^{(1)}\rangle_{\teta}
+\langle\tphi_N^{(0)}{}^{2p}w^{(1)}\rangle_{\teta}
-\langle\tphi_N^{(0)}{}^{2p}\rangle_{\teta}\langle w^{(1)}\rangle_{\teta}
\right]\nonumber\\
& & \hspace{30mm}+O(\tlmd^2)
\end{eqnarray}
that can be calculated as the B-type case before. We summarize the result without/with the weight factor in Table~\ref{table_perturb}.

\noindent
\begin{table}[htp]
\caption{Summary of first-order perturbation }
\begin{center}
\begin{tabular}{l|l|l}
\hline\hline
Type\rule[-10pt]{0pt}{28pt} & $\lim_{N\rightarrow\infty}\langle\phi_N{}^{2p}\rangle_{\eta}$ & $\lim_{N\rightarrow\infty}\langle\phi_N{}^{2p}\rangle_{\eta,w}$ \\
\hline\hline
A-type \rule[-10pt]{0pt}{28pt}\!\! & $\frac{(2p-1)!!}{m^{2p}}\left(1-\frac{\lambda}{m^4}\frac{p(4p+8+3p\,\epsilon^2m^4)}{4+\epsilon^2m^4}+O(\lambda^2)\right)$ & $\!\frac{(2p-1)!!}{m^{2p}}\left(1-\frac{\lambda}{m^4}p(p+2)+O(\lambda^2)\right)\!$\\
\hline
B-type \rule[-10pt]{0pt}{28pt} & $\frac{(2p-1)!!}{m^{2p}}\left(1-\frac{\lambda}{m^4}\frac{4p+p^2(2+\epsilon\, m^2)}{2+\epsilon\, m^2}+O(\lambda^2)\right)$ & $\!\frac{(2p-1)!!}{m^{2p}}\left(1-\frac{\lambda}{m^4}p(p+2)+O(\lambda^2)\right)\!$\\
\hline\hline
\end{tabular}
\end{center}
\label{table_perturb}
\end{table}%

\subsection{Numerical tests}
\label{num_test}
We test our proposal for stochastic quantization Eq.~\eqref{latticePW} numerically for the toy model Eq.~\eqref{0dmodel}.
The lattice drift force $W_n$ and relevant quantity $\overline{W}_n$ are not uniquely determined. 
We investigate two discrete drift forces, which we refer to as the A-type and B-type, with $\overline{W}_n=W_n$.
They are given in Eqs. \eqref{V'_ST} and \eqref{V'_CLR}, respectively, and  the associated weights are  Eqs.~\eqref{w_ST} and \eqref{w_CLR}.

The results depend on four parameters,  $\epsilon$, $\lambda$, $N$ and the number of configurations $N_s$. 
Taking $m^2$ as a unit, we use the dimensionless lattice spacing $\tilde m^2=\epsilon m^2$ and  dimensionless coupling constant $\tilde \lambda/\tilde m^4=\lambda/m^4$ as  fundamental parameters.  
We investigate two cases, $\lambda/m^4=0.01$ (weak coupling) and  $\lambda/m^4=1$ (strong coupling), 
and show the results for several lattice spacing within the range of $\tilde m^2=0.01 \sim 1$. 
Large $\tau$ means that it is sufficiently large compared to $m^2$, 
that is $m^2 \tau=\tilde m^2 N \gg 1$.  So we take $N$ to satisfy $\tilde m^2 N=10$ for each lattice spacing $\tilde m^2$.
The number of configurations is fixed as $N_s=10^6$.
The statistical error is shown as one in the 95\% CL.

The discrete Langevin equation is an implicit equation for both drift forces and is given by  the cubic equation 
on $x\equiv \tilde \phi_n$;
\begin{equation}
    a x^3+bx^2+c x +d=0,
    \label{cubic}
\end{equation}
\noindent
where $a,b,c,d$ are the real numbers presented in Table \ref{Tab2}.
Such equations can generally be solved numerically using Newton's method. 
In the present study, instead of Newton's method, we use
the Cardano formula for cubic equations to obtain the solution, 
thanks to the simplicity of the toy model. 
Eq.\eqref{cubic} has a manifestly real solution $x_1$  for $\lambda>0$ and $m^2>0$ as
\begin{eqnarray}
    x_1=  -\frac{b}{3a}+\sqrt[3]{D_+}+\sqrt[3]{D_-},   
\end{eqnarray}
where    
\begin{eqnarray}
    D_\pm\equiv \frac{-2 b^3+9 a b c -27a^2 d}{54 a^3} 
\pm \frac{\sqrt{3(27 a^2 d^2-18 a b c d+4 b^3 d+4 a c^3- b^2 c^2)}}{18a^2},
\end{eqnarray}
with $\omega=e^{2\pi i/3}$. The other two solutions $ x_2=  -\frac{b}{3a}+\omega\sqrt[3]{D_+}+\omega^2\sqrt[3]{D_-}$ and $x_3=x_2^*$
are complex numbers with nonzero imaginary parts and are irrelevant in this case. 
Therefore we adopt $x_1$ as a solution of the discrete Langevin equation. 

\begin{table}[t]
\caption{Coefficients for cubic equations setting $\phi=\tilde \phi_{n-1}$ and $\eta = \tilde \eta_n$. }
\begin{center}
\begin{tabular}{  l| c c  c c }
\hline
{}\rule[-10pt]{0pt}{24pt} & $a$ & $b$ & $c$ & $d$ \\ 
\hline
A-type \rule[-10pt]{0pt}{26pt}
& $\frac{\tilde\lambda}{2}$ & 0 & $1+\frac{\tilde m^2}{2}  $
& $ \frac{(\tilde m^2-2) }{2} \phi   
+\frac{\tilde\lambda}{2} \phi^3 
 -\eta $ \\
B-type \rule[-10pt]{0pt}{26pt}
& $\frac{\tilde\lambda}{4}$ & $\frac{\tilde\lambda}{4} \phi $ & $1+\frac{\tilde m^2}{2} 
+\frac{\tilde\lambda}{4} \phi^2$  & $
\frac{(\tilde m^2-2) }{2} \phi   +\frac{\tilde\lambda}{4} \phi^3   -\eta $  \\ 
%
\hline
\end{tabular}
\end{center}
\label{Tab2}
\end{table}

Let $\phi_n$ be the solution at time $n$ with $\phi_0=0$.
We obtain $\phi_n$ for a given noise by solving the discrete Langevin equation recursively $n$ times.  
We repeat this operation $N_s$ times changing the noise $\eta_n$ to obtain $\phi_n(i) (i=1,2,\cdots,N_s)$.
In the standard ({\it unweighted}) method, the expectation value is evaluated as a simple average over $N_s$ samples, 
\begin{eqnarray}
\langle  \phi^{2p}_N \rangle_\eta = \frac{1}{N_s} \sum_{i=1}^{N_s} (\phi_N(i))^{2p}.
\label{std_method}
\end{eqnarray}
On the other hand,  in our {\it weighted} method, the expected value is given by
\begin{eqnarray}
\langle  \phi^{2p}_N \rangle_{\eta,w} = \frac{\langle \phi^{2p}_N w \rangle_\eta}{\langle  w \rangle_\eta}, 
\end{eqnarray}
where $\langle  \cdots \rangle_\eta$ of the denominator and numerator are evaluated using Eq.\eqref{std_method}.

The validity of our method can be examined by checking the relative error 
\begin{eqnarray}
\Delta \langle  \phi^{2p}_N \rangle \equiv \frac{\langle  \phi^{2p}_N \rangle - \langle  x^{2p} \rangle_x }{\langle  x^{2p} \rangle_x}
\label{r_error}
\end{eqnarray}
as a measure of accuracy. Here $ \langle  x^{2p} \rangle_x$ is the expectation value in the toy model. 
We calculate $ \langle  x^{2p} \rangle_x$ using a numerical integration formula such as Simpson's rule:  
$m^2 \langle  x^{2} \rangle_x=0.972143853$ for $\lambda/m^4=0.01$  and $m^2 \langle  x^{2} \rangle_x=0.467919917$ for  $\lambda/m^4=1$.  
The first-order result of the perturbation theory is $m^2\langle  x^{2} \rangle^{\rm pert.}_x=1-3\lambda/m^4$ from Eq.\eqref{2pexact}.
For $\lambda/m^4=0.01$, $m^2\langle  x^{2}\rangle^{\rm pert.}_x=0.97$ is close to
the numerical result. On the other hand, $\lambda/m^4=1$ can be seen as a non-perturbative region that cannot be reproduced by perturbation calculations since $m^2\langle  x^{2} \rangle^{\rm pert.}_x=-2$.

\begin{figure}[htbp]
\begin{minipage}[b] {0.45\linewidth}
\centering
 \includegraphics[keepaspectratio,scale=0.45]{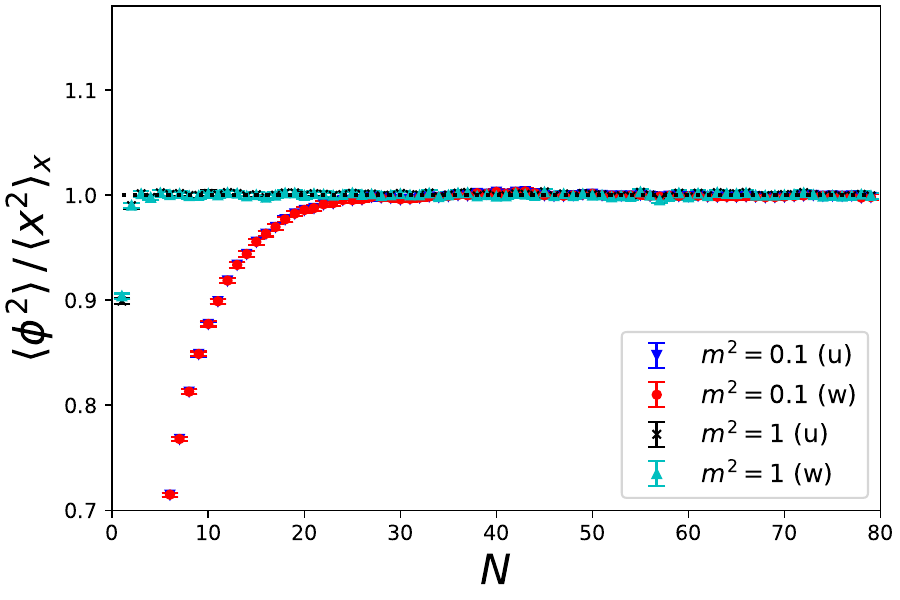}
\end{minipage}
\quad 
\begin{minipage}[b]{0.45\linewidth}
  \centering
 \includegraphics[keepaspectratio,scale=0.45]{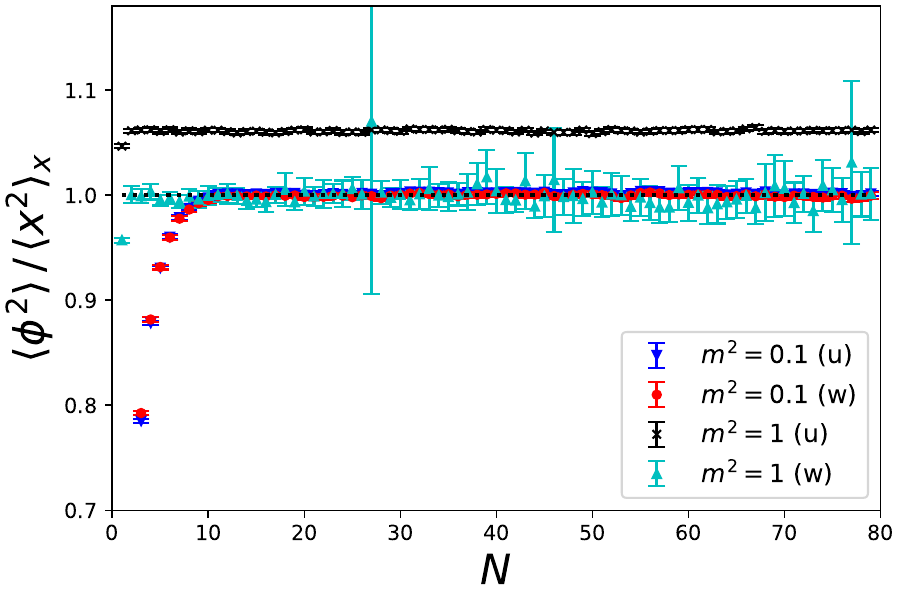}
\end{minipage}
\caption{$N$ dependence of $\langle  \phi^2_{N} \rangle$ for the A-type drift force.
The left and right figures correspond to the coupling constants $\lambda/m^4=0.01$
and $\lambda/m^4=1$, respectively. 
Each figure shows the unweighted $($u$)$ and weighted $($w$)$ results for two different lattice spacing $\tilde m^2=0.1$ and 
 $\tilde m^2=1$. The number of samples $N_s$ is fixed at $N_s=10^6$. }
\label{fig:ST1}
\vspace{2cm}
%
%
%
\begin{minipage}[b] {0.45\linewidth}
\centering
  \includegraphics[keepaspectratio,scale=0.45]{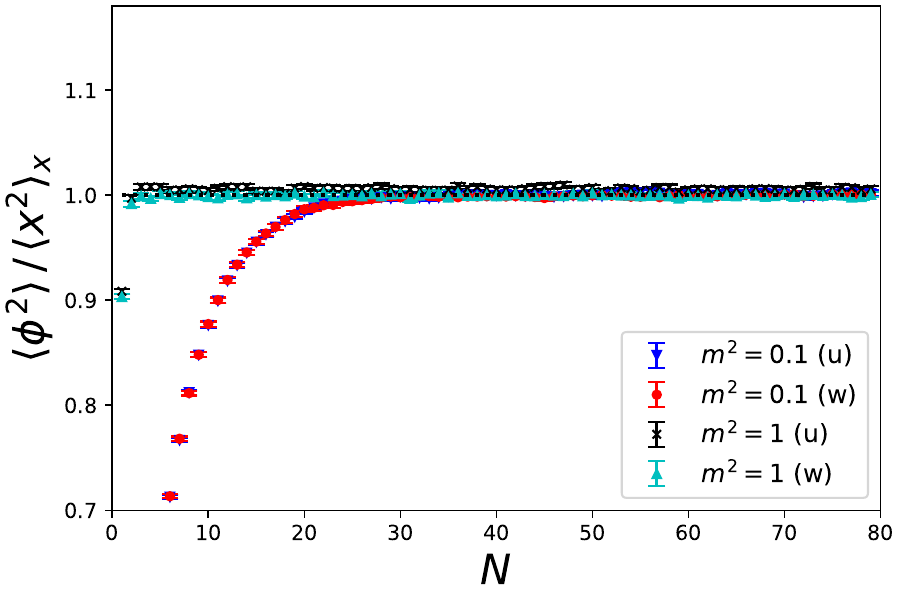}
\end{minipage}
\quad 
\begin{minipage}[b]{0.45\linewidth}
  \centering
 \includegraphics[keepaspectratio,scale=0.45]{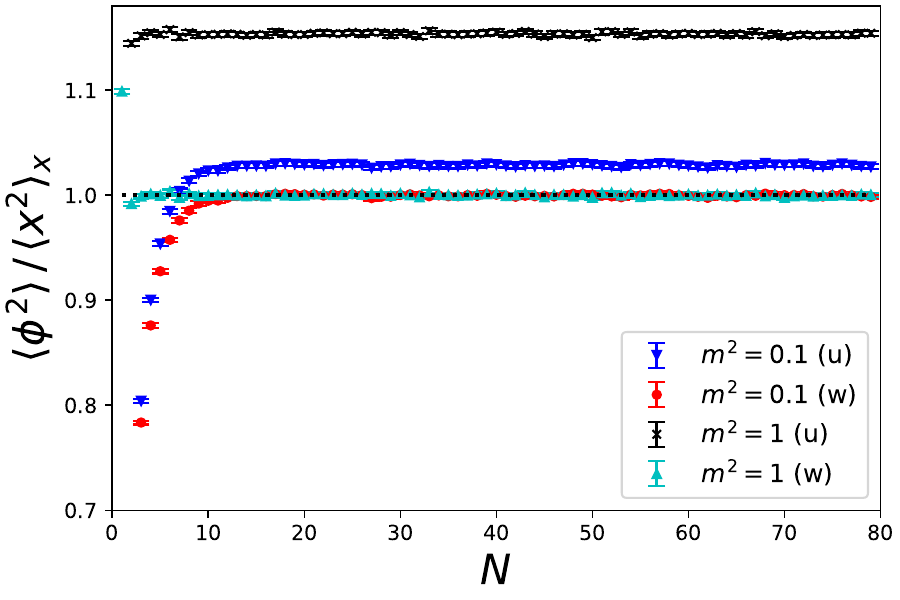}
\end{minipage}
\caption{$N$ dependence of $\langle  \phi^2_{N} \rangle$ for the B-type drift force.
The left and right figures correspond to the coupling counstants $\lambda/m^4=0.01$
and $\lambda/m^4=1$, respectively. 
Each figure shows the unweighted $($u$)$ and weighted $($w$)$ results for two different lattice spacing $\tilde m^2=0.1$ and 
 $\tilde m^2=1$. The number of samples $N_s$ is fixed at $N_s=10^6$. }
\label{fig:CLR1}
\end{figure}

Figures \ref{fig:ST1} and \ref{fig:CLR1} show $N$ dependence of $\langle  \phi^2_{N} \rangle$ for the A-type 
and the B-type drift forces, respectively. 
Since $\langle  \phi^2_{N} \rangle/\langle  x^2 \rangle_x$ is drawn on the vertical axis, 
the dotted line means that $\langle  \phi^2_{N} \rangle$ reproduces the exact solution.
The left and right figures correspond to two coupling constants  $\lambda/m^4=0.01$
and $\lambda/m^4=1$, respectively. 
Each figure shows four results with two different lattice spacings $\tilde m^2=0.1$ and $\tilde m^2= 1$
with or without weights. The $($u$)$ and $($w$)$ in Figures \ref{fig:ST1} and \ref{fig:CLR1} denote the unweighted and weighted results, respectively.

These figures show that the smaller the lattice spacing, the slower the convergence.
The convergence at large $\tau$ may be expected to behave as 
${\rm exp}(-cm^2\tau)={\rm exp}(-c\tilde m^2N)$, where $c$ is an ${\it O}(1)$ coefficient.
It can be seen that, in all cases, the results sufficiently converge for $\tilde m^2 N \sim 5$. 
Therefore, we confirm that  $\tilde m^2 N=10$, which will be used in the analysis, is large enough for the value of $N$.

In the free field limit, the weight factor becomes one. 
The two results differ little for the weak coupling. 
The effect of the weight factor is clearly visible in the strong coupling region. 
See right figures of Figures \ref{fig:ST1} and \ref{fig:CLR1}. 
Although the unweighted results deviate from the exact value of the toy model,  
the weighted ones reproduce it within the statistical errors. 

The weighted results of the A-type have large errors for the strong coupling  at $\tilde m^2=1$. 
The cause of the large errors is not entirely clear. However, the A-type weight factor contains an exponential function, which may amplify the fluctuations and increase the error. In contrast, the B-type weight does not include exponential functions, 
which may sufficiently suppress statistical fluctuations.

\begin{figure}[htbp]
    \begin{center}
 \includegraphics[width=10 cm]{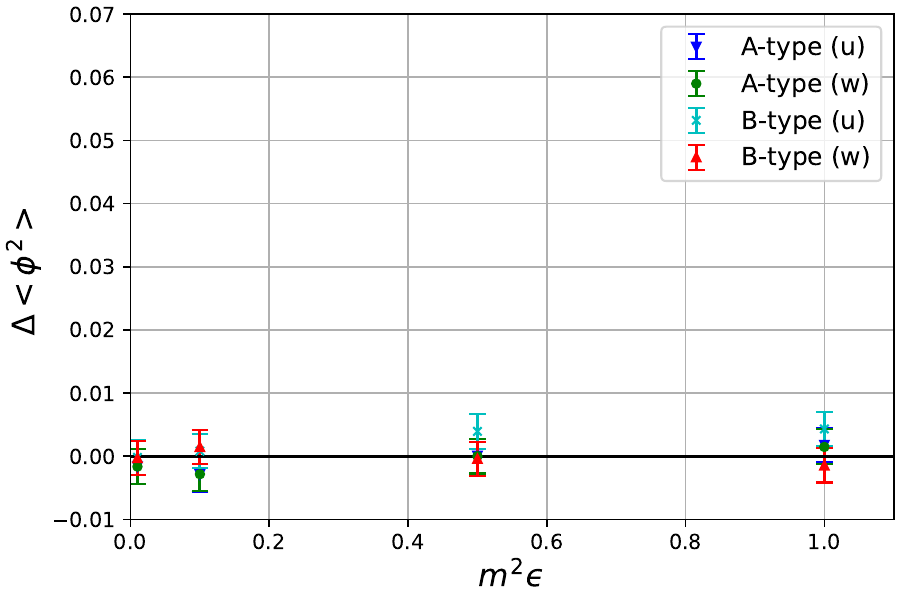}
    \end{center}
    \caption{$\epsilon$ dependence of $\Delta \langle \phi^2_N \rangle$ in $\lambda/m^4=0.01$. 
The unweighted $($u$)$ and weighted $($w$)$ results with the A and B-type drift forces are plotted as the relative error \eqref{r_error}.
$N$ is fixed so that $N\tilde m^2=10$ and $N_s$ is fixed at $N_s=10^6$. 
The weighted results coincide with the exact values for all lattice spacings used in the calculation. 
In the limit of $\lambda \rightarrow 0$, the weight is 1, so there is no significant difference between $($u$)$ and $($w$)$ in the weak coupling region. 
    }
     \label{fig:CL_lm001}
\vspace{2cm}
%
%
    \begin{center}
 \includegraphics[width=10 cm]{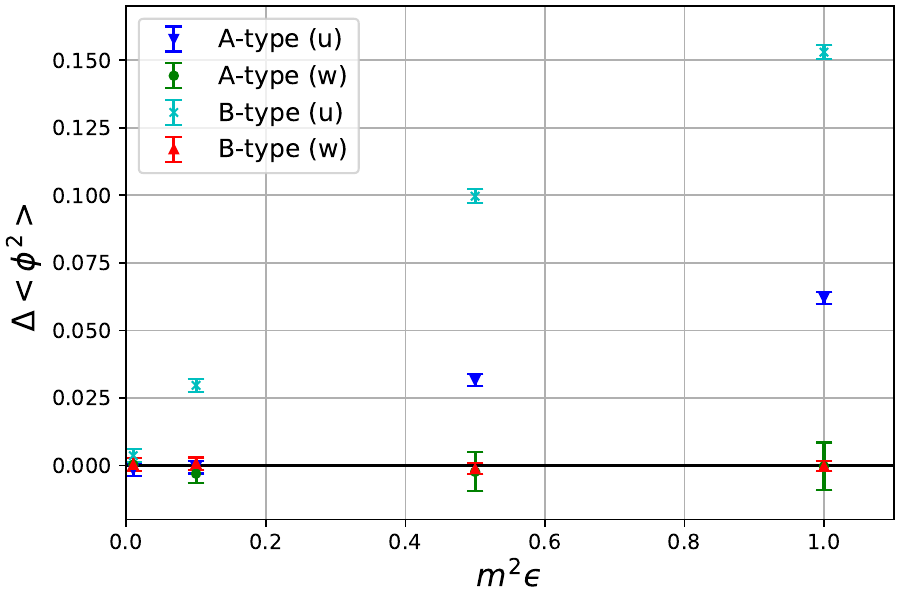}
    \end{center}
    \caption{$\epsilon$ dependence of $\Delta \langle \phi^2_N \rangle$ in $\lambda/m^4=1$. 
The unweighted $($u$)$ and weighted $($w$)$ results with the A and B-type drift forces are plotted as the relative error \eqref{r_error}.
$N$ is fixed so that $N\tilde m^2=10$ and $N_s$ is fixed at $N_s=10^6$.
Without weights, the relative error which is the difference from the exact solution vanishes in the continuous limit. On the other hand, when weights are added, the exact solution is reproduced even with nonzero lattice spacing.
    }
     \label{fig:CL_lm1}
\end{figure}

The relative error $\Delta \langle  \phi^2_N \rangle$
is plotted against the lattice spacing $\tilde m^2=m^2\epsilon$ in 
Figure~\ref{fig:CL_lm001} (weak coupling) and Figure~\ref{fig:CL_lm1} (strong coupling).
$N$ is fixed so that $\tilde m^2 N=100$ for all lattice spacings. 

In the weak coupling region (Figure~\ref{fig:CL_lm001}), 
there is no difference between weighted (red circles) and the unweighted results (blue triangles)
since $w=1$ for $\lambda=0$. 
In this case, for $\tilde m^2 \lesssim 1$,  obtained $\langle  \phi^2_N \rangle$s are almost independent on the lattice spacing, 
and both weighted and unweighted results reproduce the exact value. 

The difference between weighted and unweighted methods is clear for the strong coupling. 
As can be seen in Figure~\ref{fig:CL_lm1}, the unweighted results (blue triangles) deviate from the exact values at large lattice spacings, 
and the exact value can be reproduced in the continuum limit. 
On the other hand, weighted results (red circles) reproduce the exact value for any lattice spacing in the range  $\tilde m^2 \lesssim 1$.
In other words, our method indeed provides a stochastic quantization that reproduces the desired expected value 
even at discrete times without a continuum limit.

The numerical results  are summarized as follows. 
The weight factor is close to one in the weak coupling region, and the effect of the weights was clear in the strong coupling region.
The fluctuations of the weight factors were smaller for the B-type than for the A-type, and  the errors in the results were well controlled for the B-type. 
For both two types, the weighted results reproduce the desired expectation values without taking the continuum limit. 
This shows that our proposed method of stochastic quantization properly works in discrete time.


\section{Summary}

We have studied stochastic quantization with discrete fictitious time in $d$-dimensional scalar field theory, 
with the aim of numerical applications. We introduced a modified noise average with a weight factor and showed that stochastic quantization is possible without taking a continuum limit of fictitious time by choosing the weight appropriately.

A key ingredient of our argument is supersymmetry. In the continuum case, it is known that the noise average can be expressed as the path integral form of a $d+1$ dimensional theory using Nicolai mapping. The resulting theory has a supersymmetry consisting of two supercharges $Q$ and $\bar Q$, the latter symmetry being important for showing the Parisi-Wu stochastic quantization.

On the other hand, in the discretized time case, it is difficult to define a lattice theory with both $Q$ and $\bar Q$ symmetries. The $Q$ symmetry arises naturally in path integral form in the process of Nicolai mapping, and $\bar Q$ is broken as time is discretized. However, the $\bar Q$ symmetry is necessary for the establishment of stochastic quantization. The weight factor $w$ fills this gap. In other words, by introducing $w$ and choosing it appropriately, we can transform the $Q$-symmetric path integral into a $\bar Q$-symmetric path integral, which allows us to establish stochastic quantization in discrete time without taking a continuous limit of time.

There is some arbitrariness in the definition of the discrete Langevin equation and the weight factors. The arbitrariness is characterized by two drift terms $W$ and $\bar W$, and our argument holds if the drift terms satisfy the conditions described in Section 3. We have tested our method  both perturbatively and numerically on a zero-dimensional toy model, and found that our method works well for two types of drift forces used in the tests. 

Our next task beyond the toy model is to show the feasibility of our method in more interesting cases such as quantum mechanics and higher dimensional field theories. We need a more efficient algorithm that implements our method for such cases. We are developing a Markov chain version of our method which will be reported soon.

\section*{Acknowledgements}

HS is so much obliged to M.Suwa for her assistance in numerical analysis.
DK would like to thank Y.Kikukawa,  Y.Tanizaki  and N.Ukita for their valuable comments on our work. 
This work was supported by JSPS KAKENHI Grant Numbers JP20K03966, JP21K03537, JP22H01222, JP23K03416, JP23K22493, JP23H00112.

\appendix 
\renewcommand{\theequation}{\Alph{section}.\arabic{equation}}
\setcounter{equation}{0}
\setcounter{figure}{0}

\section{Refinement of section \ref{SQ}}
\label{A}

In this section, we present a refinement of the proof in Section \ref{SQ} 
expressing the path integral as the discretized form.
Since the treatment of $x$-space is irrelevant to the discussion here,  
we consider the $d=0$ case. For notational simplicity, we denote $f(t_n)$ as $f_n$. 

Divide the interval $[0,\tau]$ into $N$ equal parts and 
consider the discrete time $t_n=n\epsilon \, (n=0,1,\cdots,N)$
with $\epsilon = \tau /N$.   
Eq.~\eqref{eq_cor2} can be expressed as a limit 
of a discrete form:\footnote{
Here $ [d\eta] \equiv c \prod_{n=1}^N d\eta_n $ and  
$[d\phi d\bar\psi d\psi dH] \equiv c^\prime\prod_{n=1}^N d\phi_n d\bar \psi_n 
d\psi_n dH_n$ and normalization factors $c,c^\prime$ are chosen so that $\langle 1\rangle_\eta=1$. 
}
\begin{align}
\langle \phi^\ell_\eta(\tau)  \rangle_\eta 
&=\lim_{N\rightarrow \infty}
 \int [d\eta] \, 
\phi_\eta^\ell(\tau) \, 
\, {\rm exp}\left\{-\frac{\epsilon}{4} \sum_{n=1}^N \eta^2_n \right\}
\label{eq1_appA}
\\
 &= \lim_{N\rightarrow \infty}
 \int [d\phi d\bar\psi d\psi dH] \,
 \phi^\ell(\tau) \, 
 {\rm e}^{-S },
\label{eq2_appA}
\end{align}
where $S=S_1$ given by
\begin{align}
S_1 =  \epsilon \sum_{n=1}^N \, \left\{ H_n^2 + iH_n (\nabla \phi_n + L^\prime_{\textrm{QFT}}[\phi_n])
 + \bar\psi_n (\nabla  + L^{\prime\prime}_{\textrm{QFT}}[\phi_n]) \psi_n
 \right\} .
\label{susy_action_e}
\end{align}
Eq.~\eqref{eq2_appA} is obtained using the variable transformation $\eta_n =\nabla \phi_n + L^\prime_{\textrm{QFT}}[\phi_n]$ ($n=1,2,\cdots,N$) from Eq.~\eqref{eq1_appA}. 
The whole dynamical variables $\phi_n, \psi_n, \bar\psi_n, H_n$ are defined for $n=1,2,\cdots,N$
and $S$ is given with a boundary condition,  
\begin{align}
\phi_0 = \varphi_0, \qquad \psi_0=0,
\end{align}
where $\varphi_0$ is fixed. 

The action $S$ that gives the same $\langle \phi^\ell_\eta(\tau)  \rangle_\eta $ through Eq.~\eqref{eq2_appA} is not unique because
${\it O}(\epsilon)$ terms added to $S_1$ as $S=S_1+{\it O}(\epsilon)$ vanishes in the large $N$ limit ($\epsilon \rightarrow 0$). 
Using the indefiniteness of higher-order terms, the action can also be expressed as 
\begin{align}
S = S_2 + S_{\textrm{QFT}}[\varphi_\tau] - S_{\textrm{QFT}}[\varphi_0]  +{\it O}(\epsilon), 
\end{align}
where 
\begin{align}
S_2 =   \epsilon \sum_{n=1}^N \, \left\{ \bar H_n^2 - i\bar H_n (\nabla \phi_n - L^\prime_{\textrm{QFT}}[\phi_n])
 + \psi_n (\nabla  - L^{\prime\prime}_{\textrm{QFT}}[\phi_n]) \bar\psi_{n+1}
 \right\}
\label{susy_action_2}
\end{align}
with
\begin{align}
\varphi_\tau \equiv \phi_N, \qquad  
\bar\psi_{N+1}=0 .
\end{align}
Under Eq.~\eqref{free_kin}, the boundary condition was $\bar\psi_N=0$. 
This occurs when using a discretization where $\bar\psi_N$ is not included in the interaction term.
The discretization method is arbitrary, and if $\bar \psi_N$ is included in the interaction term, it is natural to set $\bar\psi_{N+1}=0$ as done in this appendix. Since $\bar\psi_{N+1}-\bar\psi_N={\it O}(\epsilon)$, in both cases the continuous counterpart becomes $\bar\psi(\tau)=0$.

Here $\bar H_n$ ($n=1,2, \cdots, N$) is   another auxiliary field introduced as
\begin{align}
\bar H_n = H_n + i\nabla \phi_n. 
\end{align}
Note that $\phi_N(=\varphi_\tau)$ is integrated in the path integral, 
and the other fields do not satisfy any boundary conditions.

The deformed theory is defined as
\begin{align}
\tilde S & \equiv S_1 + S_{\textrm{QFT}}[\varphi_0] 
\label{key relation1_appA}
\\ 
&= S_2 + S_{\textrm{QFT}}[\varphi_\tau] + {\it O}(\epsilon), 
\label{key relation2_appA}
\end{align}
changing the boundary condition at $t=0$ and considering $\varphi_0$ is a dynamical field. 

Supersymmetric transformations in discretized time are given by
\begin{alignat}{4}
&Q \phi_n      = \psi_n,  & & \bar{Q} \phi_{n-1} = \bar{\psi}_n, \\
&Q \psi_n      = 0,        & & \bar{Q} \bar \psi_n = 0,\\
&Q \bar{\psi}_n= -iH_n,  \qquad \qquad & & \bar{Q} \psi_n = i\bar H_n,  \\
&Q H_n         = 0,        & &  \bar{Q} \bar H_n = 0
\end{alignat}
for $n=1,2,\cdots, N$, and 
\begin{align}
Q \varphi_0 = 0, \ \ \ \ \bar Q \varphi_\tau=0. 
\end{align}
These transformations satisfy $Q^2=\bar Q^2=0$ for the whole fields,
and $\{Q,\bar Q\}=-\nabla +{\it O}(\epsilon) $ except  
the fields $\varphi_0$, $\varphi_\tau$, $H_N$, $\bar H_1$.

The actions $S_1$ and $S_2$ are $Q$ and $\bar Q$-invariant, respectively, 
because they can be expressed as $Q$ and $\bar Q$-exact forms without ${\it O}(\epsilon)$ terms as 
$S_1=Q(\cdots)$ and $S_2=\bar Q(\cdots)$. 
Note that from Eq.~\eqref{key relation1_appA}, the deformed action $\tilde S$ is exactly $Q$-invariant without taking $\epsilon \rightarrow 0$. 

Through some calculations, we can show that a discrete counterpart of Eq.~\eqref{key_relation} holds:
\begin{align}
\tilde S &= S_{\textrm{QFT}}[\varphi_0]  - Q\bar Q  \epsilon \sum_{n=1}^N \, I_n + {\it O}(\epsilon), \\
&= S_{\textrm{QFT}}[\varphi_\tau]  +\bar Q Q  \epsilon \sum_{n=1}^N \, I_n + {\it O}(\epsilon), 
\end{align}
where 
\begin{align}
I_n= \bar\psi_n \psi_n + S_{\textrm{QFT}}[\phi_n].
\end{align}
Thus we find that the proof from Eq.~\eqref{susy_action_offshell} to  Eq.~\eqref{key_relation} holds 
strictly in the limit $\epsilon \rightarrow 0$ from the corresponding discrete forms.

The action $\tilde S$ can also be expressed as
\begin{align}
\tilde S = S_{\textrm{QFT}}[\varphi_\tau] + 
 \bar Q\epsilon \sum_{n=1}^N \, \left\{ \psi_n ( -i\bar H_n -\nabla \phi_n + L^\prime_{\textrm{QFT}}[\phi_{n-1}] ) \right\} + {\it O}(\epsilon).
\end{align}
If we take  a discretized form of Eq.\eqref{V} as 
\begin{align}
{\cal V} = \epsilon \sum_{n=1}^N  \psi_n \nabla \phi_n, 
\end{align}
we have
\begin{align}
S_{u=1} &=  S_{\textrm{QFT}}[\varphi_\tau] + \bar Q\epsilon \sum_{n=1}^N \, \left\{ \psi_n ( -i\bar H_n + L^\prime_{\textrm{QFT}}[\phi_{n-1}] ) \right\} + {\it O}(\epsilon) \\
&=   S_{\textrm{QFT}}[\varphi_\tau]  + \epsilon \sum_{n=1}^N \, \left\{ \bar H_n^2  +i \bar H_n L^\prime_{\textrm{QFT}}[\phi_{n-1}] + \bar\psi_n L^{\prime\prime}_{\textrm{QFT}}[\phi_{n-1}] \psi_n ) \right\} + {\it O}(\epsilon).
\end{align}
We find that $\varphi_\tau(=\phi_N)$ decouples to  the bulk fields $\phi_n$ ($n \le N-1$)  in $\epsilon \rightarrow 0$ for $u=1$.
The proof using the equations from Eq.~\eqref{Su} to Eq.~\eqref{cont_dcdu} is thus justified.

\section{Perturbative expansion in a toy model}
\label{A.perturb}

Let us make a perturbative analysis in this appendix.
We start with Eq.(\ref{0dDL1})
\begin{equation}
\tphi_n = b \tphi_{n-1}+c \teta_n
-\tlmd \frac{c}{4}(\tphi_n{}^3+\tphi_n{}^2\tphi_{n-1}+\tphi_n\tphi_{n-1}{}^2+\tphi_{n-1}{}^3).
\end{equation}
We will solve this in a series of $\tlmd$ expansions:
\begin{equation}
\tphi_n=\tphi_n^{(0)}+\tlmd\tphi_n^{(1)}+\tlmd^2\tphi_n^{(2)}+\cdots.
\end{equation}
Plugging this into the above equation we obtain the following relations in each order of $\tlmd$,
\begin{eqnarray}
O(\tlmd^0):\quad\tphi_n^{(0)}&=&b\tphi_{n-1}^{(0)}+c\teta_n,\label{A0theq}\\
O(\tlmd^1):\quad\tphi_n^{(1)}&=&b\tphi_{n-1}^{(1)}-\frac{c}{4}(\tphi_n^{(0)}{}^3+\tphi_n^{(0)}{}^2\tphi_{n-1}^{(0)}+\tphi_n^{(0)}\tphi_{n-1}^{(0)}{}^2+\tphi_{n-1}^{(0)}{}^3),\label{A1steq}\\
&\vdots&\nonumber
\end{eqnarray}
We start with the simplest initial condition $\tphi_0=0$ since any other choice does not affect the final result at $N\rightarrow\infty$. Also, we assume $\tms> 0$ ({\it i.e.} $|b|< 1$).

The 0-th order equation (\ref{A0theq}) can be solved immediately
\begin{equation}
\tphi_n^{(0)}=c\sum_{l=1}^nb^{n-l}\teta_l.
\end{equation}
Then the 1-st order equation (\ref{A1steq}) leads to
\begin{eqnarray}
\tphi_n^{(1)}&=&-\frac{c}{4}\sum_{l=1}^nb^{n-l}
(\tphi_l^{(0)}{}^3+\tphi_l^{(0)}{}^2\tphi_{l-1}^{(0)}+\tphi_l^{(0)}\tphi_{l-1}^{(0)}{}^2+\tphi_{l-1}^{(0)}{}^3) \nonumber\\
&=&-\frac{c^4}{4}\left[
\sum_{l=1}^n\sum_{l_1=1}^l\sum_{l_2=1}^l\sum_{l_3=1}^lb^{n-l+\sum_{i=1}^3(l-l_i)}\teta_{l_1}\teta_{l_2}\teta_{l_3}\right.\nonumber\\
& &\hspace{7mm}+\sum_{l=2}^n\sum_{l_1=1}^l\sum_{l_2=1}^l\sum_{l_3=1}^{l-1}b^{n-l+\sum_{i=1}^3(l-l_i)-1}\teta_{l_1}\teta_{l_2}\teta_{l_3}\nonumber\\
& &\hspace{7mm}+\sum_{l=2}^n\sum_{l_1=1}^l\sum_{l_2=1}^{l-1}\sum_{l_3=1}^{l-1}b^{n-l+\sum_{i=1}^3(l-l_i)-2}\teta_{l_1}\teta_{l_2}\teta_{l_3}\nonumber\\
& &\hspace{7mm}\left.+\sum_{l=2}^n\sum_{l_1=1}^{l-1}\sum_{l_2=1}^{l-1}\sum_{l_3=1}^{l-1}b^{n-l+\sum_{i=1}^3(l-l_i)-3}
\teta_{l_1}\teta_{l_2}\teta_{l_3}\right].
\end{eqnarray}

Using these results we can evaluate the $2p$-point function without weights
\begin{eqnarray}
\langle\tphi_N{}^{2p}\rangle_{\teta}
&=&\langle\tphi_N^{(0)}{}^{2p}\rangle_{\teta}
+2p\tlmd\langle\tphi_N^{(0)}{}^{2p-1}\tphi_N^{(1)}\rangle_{\teta}
+O(\tlmd^2)\nonumber \\[10pt]
&=&(2p-1)!!(2c^2)^pH(b^2,N)^p\nonumber\\
& &-(2p-1)!!c^{2p+3}2^pp\tlmd
\left[\rule{0pt}{15pt}\right.\nonumber\\
& &H(b^2,N)^{p-1}\sum_{l=1}^Nb^{2(N-l)}
\left\{3H(b^2,l)^2+(3b+1)H(b^2,l)H(b^2,l-1)\right.\nonumber\\
& &\hspace{40mm}\left.+(2b^2+3b)H(b^2,l-1)^2\right\}\nonumber\\
& &\hspace{1mm}+(2p-2)H(b^2,N)^{p-2}\sum_{l=1}^Nb^{4(N-l)}
\left\{H(b^2,l)^3+bH(b^2,l)^2H(b^2,l-1)\right.\nonumber\\
& &\hspace{25mm}\left.\left.+b^2H(b^2,l)H(b^2,l-1)^2+b^3H(b^2,l-1)^3\right\}\rule{0pt}{15pt}\right]+O(\tlmd^2),\qquad
\end{eqnarray}
where
\begin{equation}
H(x,n)\equiv\frac{1-x^{n}}{1-x}.
\end{equation}
Taking a limit $N\rightarrow\infty$ we obtain
\begin{equation}
\tms{}^p\langle\tphi_N{}^{2p}\rangle_{\teta}^{\rm B} \underset{N\rightarrow\infty}{\longrightarrow}(2p-1)!!\left(1-\frac{\tlmd}{\tilde m^4}\frac{4p+p^2(2+\tms)}{2+\tms}+O(\tlmd^2)\right).\label{A2pCLR}
\end{equation}
The right-hand side depends not only $\frac{\tlmd}{\tilde m^4}$ but also $\tms$, so we need the continuum limit $\tms\rightarrow 0$ with $\frac{\tlmd}{\tilde m^4}$ fixed to reproduce the perturbative result of the 0-dimensional integral
\begin{equation}
\langle x^{2p}\rangle_x = \frac{(2p-1)!!}{m^{2p}}\left[1-\frac{\lambda}{m^4}p(p+2)+O(\lambda^2)\right].\label{A2pexact}
\end{equation}

Now we turn to include the weight factor.
The weight factor $w$ for the B-type can  be expanded concerning the $\tlmd$ as
\begin{equation}
w^{\rm B} = 1-\frac{\tlmd}{2+\tms}\tphi_N^{(0)}{}^2 +O(\tlmd^2).
\end{equation}
Making use of this, we have 
\begin{eqnarray}
\langle\tphi_N{}^{2p}\rangle_{\teta,w}&=&\frac{\langle\tphi_N{}^{2p}w\rangle_{\teta}}{\langle w\rangle_{\teta}} \nonumber\\
&=&\langle\tphi_N^{(0)}{}^{2p}\rangle_{\teta}
+\tlmd\left[2p\langle\tphi_N^{(0)}{}^{2p-1}\tphi_N^{(1)}\rangle_{\teta}
-\frac{1}{2+\tms}\left(\langle\tphi_N^{(0)}{}^{2p+2}\rangle_{\teta}-\langle\tphi_N^{(0)}{}^{2p}\rangle_{\teta}\langle\tphi_N^{(0)}{}^2\rangle_{\teta}
\right)\right]\nonumber\\
& & \hspace{30mm}+O(\tlmd^2).
\end{eqnarray}
The correction term coming from the weight factor is evaluated as
\begin{eqnarray}
-\frac{\tlmd}{2+\tms}\left(\langle\tphi_n^{(0)}{}^{2p+2}\rangle_{\teta}-\langle\tphi_N^{(0)}{}^{2p}\rangle_{\teta}\langle\tphi_N^{(0)}{}^2\rangle_{\teta}
\right)&=&-\frac{\tlmd}{2+\tms}\left(2c^2H(b^2,N)\right)^{p+1}2p(2p-1)!!
\nonumber\\
&\underset{N\rightarrow\infty}{\longrightarrow}&-\frac{\tlmd}{2+\tms}\frac{2p(2p-1)!!}{\tilde m^{2p+2}}.
\end{eqnarray}
Combining Eq.(\ref{A2pCLR}), this leads us to the result
\begin{equation}
\langle\tphi_N{}^{2p}\rangle_{\teta,w}^{\rm B} \underset{N\rightarrow\infty}{\longrightarrow}\frac{(2p-1)!!}{\tilde m^{2p}}\left(1-\frac{\tlmd}{\tilde m^4}p(p+2)+O(\tlmd^2)\right).
\end{equation}
Alternatively, for the dimensionful quantities, we have
\begin{equation}
\langle\phi_N{}^{2p}\rangle_{\eta,w}^{\rm B} \underset{N\rightarrow\infty}{\longrightarrow}\frac{(2p-1)!!}{m^{2p}}\left(1-\frac{\lambda}{m^4}p(p+2)+O(\lambda^2)\right),
\end{equation}
which exactly coincides with the desired result (\ref{A2pexact}) up to the first order of $\lambda$ without taking any continuum limit.

\bibliographystyle{JHEP}

\bibliography{Refs}

\end{document}